\documentclass[11pt]{article}
\usepackage{wrapfig}
\usepackage{epsfig}
\usepackage{epstopdf}
\usepackage{comment,bm,dsfont}
\usepackage[colorlinks,linkcolor=blue,citecolor=blue,urlcolor=blue]{hyperref}
\usepackage{soul}
\usepackage{subcaption}
\usepackage{float}

\usepackage{setspace}

\usepackage{amsmath, amsthm, amsfonts, bbm, amssymb}
\usepackage{ifpdf}
\usepackage{graphicx}
\usepackage{wrapfig}
\usepackage[authoryear]{natbib}

\usepackage{amsmath}

\DeclareMathOperator*{\argmin}{arg\,min}
\usepackage[dvipsnames]{xcolor}
\usepackage{lipsum}%
\usepackage{sectsty}
\usepackage{wrapfig}
\def\mcZ{\mathcal{Z}}
\usepackage{tikz}
\def\spacingset#1{\renewcommand{\baselinestretch}%
{#1}\small\normalsize} \spacingset{1}
\usepackage{enumitem}
\setitemize{leftmargin=0in}
\usepackage{titlesec}
\usepackage[T1]{fontenc}
\usetikzlibrary{positioning}
\titleformat{\section}
{\normalfont\large\bfseries\centering}{\thesection}{0.5em}{}
\titleformat{\subsection}
{\normalfont\bfseries}{\thesubsection}{0.5em}{}
\titleformat{\subsubsection}
{\normalfont\normalsize\bfseries}{\thesubsubsection}{0.5em}{}
\usepackage{tcolorbox}
\titleformat{\paragraph}[runin]
{\normalfont\normalsize\bfseries}{\theparagraph}{0.5em}{}

\titleformat{\subparagraph}[runin]
{\normalfont\normalsize\bfseries}{\thesubparagraph}{0.5em}{}
\usepackage[utf8]{inputenc}

\titlespacing*{\section}{0pt}{0.3em}{0.1em}
\titlespacing*{\subsection}{0pt}{0.3em}{0.2em}
\titlespacing*{\subsubsection}{0pt}{*0}{*0}
\usepackage{enumitem}
\setlist{nosep}

\usepackage[top=1in, bottom=1in, left=1in, right=1in]{geometry}

\newtheorem{lemma}{\sc Lemma}
\theoremstyle{plain}
\newtheorem{theorem}{\sc Theorem}

\newtheorem{proposition}[theorem]{\sc Proposition}
\theoremstyle{remark}

\theoremstyle{definition}

\newtheorem{assumption}{\sc Assumption}

\def \mcH{\mathcal{H}}
\def \mcP{\mathcal{P}}
\def \mcQ{\mathcal{Q}}
\def \mcE{\mathcal{E}}
\def \mcC{\mathcal{C}}
\def \cH{\mathcal{H}}

\def \cM{\mathcal{M}}

\def \cX{\mathcal{X}}

\def \mcL{\mathcal{L}}

\def \E{\mathbb{E}}

\usepackage{bbm}

\newcommand{\mbbR}{\mathbb{R}}
\newcommand{\mbbP}{\mathbb{P}}

\newcommand{\bean}{\begin{eqnarray*}}
\newcommand{\eean}{\end{eqnarray*}}
\newcommand{\bea}{\begin{eqnarray}}
\newcommand{\eea}{\end{eqnarray}}
\newcommand{\ee}{\end{eqnarray}}
\newcommand{\beq}{\begin{equation}}

\newcommand{\eeq}{\end{equation}}

\usepackage{pdfpages}
\usepackage{pgfplots}
\usepackage{enumitem}
\setlist[itemize]{leftmargin=18mm}
\title{Nonparametric In-Context Learning and Inference via Measure-Indexed Kernel Ridge Regression}

\author{%
Zuofeng Shang\\
\small Department of Mathematical Sciences, New Jersey Institute of Technology
\\[1em]
Tianyang Hu\\
\small School of Data Science, The Chinese University of Hong Kong, Shenzhen
\\[1em]
Fan Yang\\
Eli Lilly and Company
\\[1em]
Yingcong Li 
\\
\small Department of Data Science, New Jersey Institute of Technology
}

\date{}

\begin{document}
\maketitle

\begin{abstract}
In-Context learning enables an estimator, trained across multiple tasks, to adapt to a new unseen task solely through its in-context samples, without any parameter refitting. Despite growing interest in the statistical foundations of in-context learning, two gaps still remain. First, existing convergence results do not characterize the distributional behavior of the in-context learning estimators. Second, classical nonparametric methods are not designed to handle the measure-valued inputs that in-context learning naturally produces. We address both gaps by developing a measure-indexed reproducing kernel Hilbert space framework, in which the adaptivity of the estimator is captured through the dependence of a mean functional on the empirical measure of the context. Within this framework, we propose a measure-indexed kernel ridge regression estimator computed from multiple independent tasks and establish its minimax-optimal convergence rate. We further prove that the proposed estimator is pointwise asymptotically normal, enabling the construction of confidence intervals for the mean and prediction intervals for the response on an unseen task. Notably, the estimator achieves valid uncertainty quantification without any refitting when adapting to a new task. As a special case, our framework recovers classical distributed nonparametric regression when all tasks share the same underlying function. The proposed framework is illustrated through a latent Gaussian process example.
\end{abstract}

\section{Introduction}
In-Context learning (ICL) refers to an emerging prediction paradigm in which an estimator, trained once across a collection of tasks, automatically adapts its predictions to an unseen task solely through the task's in-context samples, without any refitting of model parameters \citep{brown2020language,dong2022survey}. This contrasts sharply with conventional statistical learning, where a separate model must be retrained for each new task.
The adaptivity of ICL calls for a theoretical explanation, for which the Bayesian perspective provides a compelling account \citep{xie2022explanation};
the optimal predictor is the posterior predictive mean which is a
fixed functional of the in-context samples and the query that automatically
adapts to any new task through the context alone.
This perspective has motivated a growing body of work connecting ICL to classical statistical estimators. For linear function classes, a single layer of linear attention can implement ridge regression on the in-context samples \citep{akyurek2022learning,vonoswald2023transformers,bai2023transformers}. For nonparametric regression over smooth function classes, \citet{ching2026efficient} showed that transformers trained via meta-learning can attain minimax-optimal prediction rates, and \citet{wakayama2026incontext} established a rigorous connection between ICL and Bayesian inference for characterizing the optimal predictor and its generalization error.

Despite this rapid progress, two gaps remain. The first
is concerned with uncertainty quantification. Existing convergence results
establish that the ICL predictor converges to the truth, but say nothing
about its distributional behavior, though the adaptivity of ICL is most
valuable precisely when one needs to quantify uncertainty about a
prediction made on a previously unseen task. Constructing confidence or prediction intervals 
in this setting requires asymptotic normality, which no existing result
provides. The second gap is concerned with the geometry of the input domain. Directly using the raw context sample as input can lead to the ``curse of dimensionality,'' motivating a dimension-reduction strategy that transforms the input space into a more structured representation. The natural
mathematical object representing an in-context sample is its empirical
distribution \citep{Liu2026UnifiedFramework}. The ICL
predictor is therefore a function whose argument includes a probability
measure as part of its input. Classical nonparametric methods, including
the reproducing kernel Hilbert space (RKHS) based methods such as kernel ridge regression or
related ones, are mostly built for Euclidean domains. They do not directly accommodate this measure-valued structure. Adapting
them requires a nontrivial generalization of the standard RKHS framework.

In this paper, we propose a new RKHS framework, called measure-indexed RKHS (MI-RKHS), 
with measure-valued inputs 
for ICL that addresses both
gaps simultaneously. The key observation is that ICL defines a fixed
mean functional from a (context, query) pair to predictions; 
the adaptivity of the estimator across tasks is
naturally captured by the dependence of this functional on the empirical
distribution of the context. We model this 
function as an element in an MI-RKHS 
defined over the product of a space of probability measures and a feature
space, which allows classical RKHS analysis to be extended to the
measure-valued setting. For instance, we propose a measure-indexed kernel ridge
      regression (MI-KRR) estimator computed from multiple
      independent tasks and establish its convergence rate. 
We prove that the MI-KRR
      estimator is pointwise asymptotically normal, enabling the
      construction of confidence intervals for the mean and prediction intervals for the response on an unseen task. Notably, the estimator adapts to a new task without any refitting, yet still achieves asymptotic normality.
      The proposed MI-KRR method extends classical kernel ridge regression
to the measure-valued setting, with both convergence rate and asymptotic
normality established within this new framework.
As a special case, when all tasks share the same fixed underlying function,
our framework degenerates to the well-known distributed nonparametric regression,
recovering known results in that literature such as optimal convergence rate \citep{zhang2015divide,shang2017jmlr,shang2019jmlr,shang:ejs:2020,szabo2020aos}.
More generally, our framework is conceptually related to meta-learning or transfer learning 
since both paradigms require the underlying tasks are different but still possess similarity.

The rest of this paper is organized as follows. In Section \ref{sec:hyper-RKHS},
we introduce the MI-RKHS framework as well as a MI-KRR regression for estimating the mean functional. In Section \ref{sec:asymptotics}, we derive the convergence rate and asymptotic distribution for the proposed MI-KRR estimator. An example on latent Gaussian process model is provided to examine the proposed MI-RKHS framework. Proofs of the theorems and lemmas are deferred to Section \ref{sec:appendix}. 

\section{In-Context Learning in Measure-Indexed RKHS Framework}\label{sec:hyper-RKHS}
Let $\cX\subseteq\mathbb{R}^d$ be a $d$-dimensional feature space, and $\cH_X$ be an RKHS of functions on $\cX$
endowed with inner product $\langle\cdot,\cdot\rangle_{\cH_X}$ and reproducing kernel $R^{\langle X\rangle}:\cX\times\cX\to\mbbR$.
Suppose that $\mu$ is a known deterministic probability measure on $\cX$. We further assume that $R^{\langle X\rangle}$ is bounded.
\begin{assumption}\label{A0} 
$C_X\triangleq\sup_{x\in\cX}R^{\langle X\rangle}(x,x)<\infty$.
\end{assumption}
The Assumption \ref{A0} is standard in the nonparametric literature which is satisfied for commonly used kernels such as Sobolev kernel and Gaussian kernels \citep{shang2013aos,shang2020colt}. 
Let $f$ be a latent random function with distribution $F$, where $F$ is supported on $\mathcal H_X$. Suppose $X_1,\ldots,X_{n+1}$ are i.i.d. feature variables from $\mu$ and independent of $f$. The response variables $Y_i$ satisfy 
\begin{equation}\label{nonpara:model}
Y_i=f(X_i)+\varepsilon_i, \quad i=1,\ldots,n+1,
\end{equation}
where $\varepsilon_i$'s are i.i.d. with mean zero and 
finite variance, independent of
$f,X_1,\ldots,X_{n+1}$.
We are interested in estimating 
the following conditional mean functional 
\begin{equation}\label{eqn:M_*}
M_*(D_n,x)=\E[Y_{n+1}|D_n,X_{n+1}=x],\,\,\,\,x\in\cX,   
\end{equation}
where $D_n=\{Z_1,\ldots,Z_n\}$ with $Z_i=(X_i,Y_i)$.
Estimation of $M_*$ is closely related to ICL.
In classic ICL literature, $D_n$ is regarded as the context data and 
$Z_{n+1}$ is the new query pair  \citep{garg2022what,vonoswald2023transformers,Zhang2024TrainedTransformersLearnLinear,Oko2024PretrainedTransformer,Kim2024TransformersAreMinimaxOptimal,agarwal2024manyshot,vladymyrov2024linear,Liu2025InContextLearningAsNonparametric,wakayama2026incontext,ching2026efficient}. The target is to estimate $M_*$ which will be useful to predict the next query output $Y_{n+1}$ based on $D_n$ and $X_{n+1}$. 
Directly estimating $M_*$ generally suffers from the ``curse-of-dimensionality'' since the input $D_n$ is $n(d+1)$-dimensional which is high as $n$ diverges. A dimension-reduction strategy is usually needed to make this estimation problem well posed. 

An observation is that, although the ambient dimension of
$D_n$ increases with $n$, the estimation target $M_*(D_n,x)$ may admit
a substantially simpler representation. In particular, under the
conditional i.i.d. structure of $D_n$, $M_*$ is invariant to
the ordering of the observations in $D_n$ and hence depends on the
$D_n$ only through its empirical measure. This observation is
formalized as follows.
\begin{lemma}\label{lemma:dimension:reduction}
Under the conditional i.i.d. assumption on
$Z_1,\ldots,Z_{n+1}$ given $f$, it holds that
\[
\E[Y_{n+1}| D_n,X_{n+1}]
=
\E[Y_{n+1}| \widehat{\pi}_n,X_{n+1}]
\quad\text{a.s.},
\]
where
$\widehat{\pi}_n=n^{-1}\sum_{i=1}^n\delta_{Z_i}$
is the empirical measure of $Z_1,\ldots,Z_n$.
\end{lemma}
Lemma \ref{lemma:dimension:reduction} shows that, for the purpose of
predicting $Y_{n+1}$, the ordered $D_n$ can be replaced by its
empirical measure $\widehat{\pi}_n$. Thus, although $D_n$ is a
high-dimensional ordered object, $M_*$ is invariant
to permutations of the observations and can be represented through the
measure-valued input $\widehat{\pi}_n$ together with the query point
$x$. Accordingly, with a slight abuse of notation, we henceforth
regard $M_*$ as a function of $(\widehat{\pi}_n,x)$ and write
\[
M_*(\widehat{\pi}_n,x) \equiv M_*(D_n,x).
\]
Therefore, $M_*$ is a function defined on the
domain $\widehat{\mcP}(\mcZ)\times\cX$, where 
\[
\widehat{\mcP}(\mcZ)=\left\{n^{-1}\sum_{i=1}^n\delta_{z_i}: z_1,\ldots,z_n\in\mcZ, n\ge1\right\}
\]
is the space of empirical measures and $\mcZ=\cX\times\mbbR$ is the range of feature-response pairs. This allows us to
characterize the regularity of $M_*$ with respect to $\widehat{\pi}_n$, and hence, reduce the complexity of $M_*$.
This measure-valued representation is related to the recent ICL framework of
\citet{Liu2026UnifiedFramework}, where empirical measures are used as a
permutation-invariant representation of the context data to facilitate a unified
analysis across different context lengths and transformer architectures.
In contrast, Lemma~\ref{lemma:dimension:reduction} shows that the use of
$\widehat{\pi}_n$ provides an alternative representation of the conditional mean functional $M_*$ through information-preserving reduction of $D_n$,
which will facilitate our statistical analysis.

\subsection{A General MI-RKHS Framework}
In this section, we establish a general MI-RKHS framework that characterizes the regularity of $M_*$. 
Recall that $f$ is the unobservable latent function drawn from $F$, and therefore $f\in\cH_X$ almost surely. 
We further assume that $\|f\|_{\cH_X}$ has a finite second moment.
\begin{assumption}\label{A1}
$\E[\|f\|_{\cH_X}^2]<\infty$.
\end{assumption}
Assumption \ref{A1} says that $f$ has a finite second moment in terms of the $\|\cdot\|_{\cH_X}$ norm.
It ensures that the complexity of $f$ is well controlled in expectation.
By (\ref{nonpara:model}), (\ref{eqn:M_*}) and independence between $X_{n+1}$ and $(f,D_n)$, we have 
\begin{equation}\label{alt:expression:M_*}
M_*(\widehat{\pi}_n,x)=\E[f(X_{n+1})+\varepsilon_{n+1}|\widehat{\pi}_n,X_{n+1}=x]=\E[f|\widehat{\pi}_n](x).
\end{equation}
Let $H_*(\widehat{\pi}_n)=\E[f|\widehat{\pi}_n]$.
By Assumption \ref{A1}, we have
\[
\E[\|H_*(\widehat{\pi}_n)\|_{\cH_X}^2]\le\E[\|f\|_{\cH_X}^2]<\infty,
\]
so $H_*(\widehat{\pi}_n)\in\cH_X$ almost surely, implying that $H_*$ is a mapping from $\widehat{\mcP}(\mcZ)$ to $\cH_X$.
To characterize the regularity of $H_*$ over a richer domain, let 
\[
\mcP(\mcZ)=\left\{\pi:
\text{$\pi$ is a probability
measure on $\mcZ$ with }
\int_\mcZ y^2\,d\pi(x,y)
<\infty\right\}.
\]
Any $\pi\in\mcP(\mcZ)$ satisfies the condition $\int_\mathcal{Z} y^2\, d\pi(x,y) < \infty$,
which requires that the probability measure $\pi$ on $\mathcal{Z}$ has a finite second moment in the response variable $y$.
Note that $\widehat{\mcP}(\mcZ)\subset\mcP(\mcZ)$. We assume that $H_*:\widehat{\mcP}(\mcZ)\to\cH_X$ admits an extension to $\mcP(\mcZ)$, which we continue to denote by $H_*$ by a slight abuse of notation. To characterize the regularity of this extension, we assume $H_*\in\cH_\Pi$, where $\cH_\Pi$ is a vector-valued RKHS of mappings from $\mcP(\mcZ)$ into $\cH_X$. Specifically, $\cH_\Pi$ is endowed with inner product $\langle\cdot,\cdot\rangle_{\cH_\Pi}$ and operator-valued reproducing kernel $R^{\langle\Pi\rangle}:\mcP(\mcZ)\times\mcP(\mcZ)\to\mcL(\cH_X)$, with $\mcL(\cH_X)$ denoting the space of bounded linear operators on $\cH_X$. For any $\pi\in\mcP(\mcZ)$, define $R^{\langle\Pi\rangle}_{\pi}(\cdot)=R^{\langle\Pi\rangle}(\cdot,\pi)$, so $R^{\langle\Pi\rangle}_{\pi}$ is a well-defined operator from $\mcP(\mcZ)$ to $\mcL(\mcH_X)$.
Moreover, for any $g\in\cH_X$, define $(R^{\langle\Pi\rangle}_{\pi} g)(\cdot)\triangleq R^{\langle\Pi\rangle}(\cdot,\pi)g\in\cH_X$ as the kernel section at $\pi$ in the direction of $g$. Obviously, $R^{\langle\Pi\rangle}_{\pi} g$ is a mapping from $\mcP(\mcZ)$ to $\cH_X$. 
The reproducing property in $\cH_\Pi$ is stated as follows. For any $H\in\cH_\Pi$, $g\in\cH_X$ and $\pi\in\mcP(\mcZ)$,
\begin{enumerate}[label=Property~(\alph*):,ref=(\alph*), leftmargin=*]
\item\label{eqn:reproducing:property:i} $R^{\langle\Pi\rangle}_{\pi} g\in\cH_\Pi$;
\item\label{eqn:reproducing:property:ii} $\langle H(\pi),g\rangle_{\cH_X}=\langle H,R^{\langle\Pi\rangle}_{\pi} g\rangle_{\cH_{\Pi}}$.
\end{enumerate}
The object $R^{\langle\Pi\rangle}_{\pi} g$ can be interpreted as the kernel section of the operator-valued kernel $R^{\langle\Pi\rangle}$ at $\pi$ in the direction $g\in\cH_X$.
Property~\ref{eqn:reproducing:property:i} requires that $R^{\langle\Pi\rangle}_{\pi} g$ be an element of $\cH_\Pi$. Property~\ref{eqn:reproducing:property:ii} states that the evaluation of the $\cH_X$-valued function $H$ at $\pi$, when projected onto any direction $g\in\cH_X$, can be recovered as an inner product in $\cH_\Pi$. These properties are the natural operator-valued analogue of the classical RKHS reproducing property. See \citep{micchelli2005learning,alvarez2012kernels} for further discussions.

We will use the above RKHS's $\cH_X$ and $\cH_\Pi$ to construct an MI-RKHS $\cM$ of functions on $\mcP(\mcZ)\times\cX$ that contains the true function $M_*$. Our approach involves construction of kernels on $\mcP(\mcZ)\times\cX$ which is closely related to \textit{hyper-kernels} defined on distributions \citep{evgeniou2005learning, muandet2017}. 
For $H\in\cH_\Pi$ and $(\pi,x)\in\mcP(\mcZ)\times\cX$, define
$M_H(\pi,x)=\langle H(\pi),R^{\langle X\rangle}_x\rangle_{\cH_X}$. Define
\begin{equation}\label{rkhs:M}
\cM=\left\{M_H: H\in\cH_\Pi\right\},
\end{equation}
and inner product $\langle M_H, M_{H'}\rangle_\cM\triangleq\langle H,H'\rangle_{\cH_\Pi}$ for $H,H'\in\cH_\Pi$. 
Since $H_*\in\cH_\Pi$, we have
\begin{equation}\label{eqn:H_*}
M_*(\pi,x)=\langle H_*(\pi),R^{\langle X\rangle}_x\rangle_{\cH_X}=M_{H_*}(\pi,x).
\end{equation}
So $M_*=M_{H_*}\in\cM$. 
Clearly, $\cM$ is a Hilbert space since the mapping $H\mapsto M_H$ is an isometric isomorphism between $\cH_\Pi$ and $\cM$.
Let
\begin{equation}\label{reproducing:kernel:M}
\textrm{$R((\pi,x),(\pi',x'))=\langle R^{\langle\Pi\rangle}_{\pi} R^{\langle X\rangle}_x,R^{\langle\Pi\rangle}_{\pi'}R^{\langle X\rangle}_{x'}\rangle_{\cH_\Pi}$ for $(\pi,x), (\pi',x')\in\mcP(\mcZ)\times\cX$.}
\end{equation}
Since $R^{\langle X\rangle}_x,R^{\langle X\rangle}_{x'}\in\cH_X$, it follows from Property~\ref{eqn:reproducing:property:i} that both $R^{\langle\Pi\rangle}_{\pi} R^{\langle X\rangle}_x$ and $R^{\langle\Pi\rangle}_{\pi'}R^{\langle X\rangle}_{x'}$ belong to $\cH_\Pi$,
so definition (\ref{reproducing:kernel:M}) is meaningful. 
Define $R_{(\pi,x)}(\cdot,\cdot)=R((\cdot,\cdot),(\pi,x))$, then $R_{(\pi,x)}\in\cM$. 
To see this, let $H_{\pi,x}=R^{\langle\Pi\rangle}_{\pi} R^{\langle X\rangle}_x$ which is an element of $\cH_\Pi$ thanks to Property~\ref{eqn:reproducing:property:i}. It follows by Property~\ref{eqn:reproducing:property:ii} that
\begin{eqnarray*}
M_{H_{\pi,x}}(\pi',x')&=&\langle H_{\pi,x}(\pi'),R^{\langle X\rangle}_{x'}\rangle_{\cH_X}=\langle H_{\pi,x},R^{\langle\Pi\rangle}_{\pi'}R^{\langle X\rangle}_{x'}\rangle_{\cH_\Pi}\\
&=&\langle R^{\langle\Pi\rangle}_{\pi} R^{\langle X\rangle}_x,R^{\langle\Pi\rangle}_{\pi'}R^{\langle X\rangle}_{x'}\rangle_{\cH_\Pi}=R_{(\pi,x)}(\pi',x'),
\end{eqnarray*}
implying that $R_{(\pi,x)}=M_{H_{\pi,x}}\in\cM$. Moreover, the reproducing property follows from 
Property~\ref{eqn:reproducing:property:ii}. Specifically, for any $M_H\in\cM$,
\begin{equation}\label{reproducing:property:R}
\langle M_H,R_{(\pi,x)}\rangle_{\cM}=
\langle H,R^{\langle\Pi\rangle}_{\pi} R^{\langle X\rangle}_x\rangle_{\cH_\Pi}=
\langle H(\pi),R^{\langle X\rangle}_x\rangle_{\cH_X}=
M_H(\pi,x).
\end{equation}
In summarizing the above, we have proved the following result.
\begin{proposition}
$\cM$ is an RKHS of functions on $\mcP(\mcZ)$ endowed with inner product $\langle\cdot,\cdot\rangle_\cM$ and reproducing kernel $R$.
The function $M_*$ defined in (\ref{eqn:M_*}) is an element of $\cM$.
\end{proposition}

\subsection{Measure-Indexed Kernel Ridge Regression}\label{sec:loo:krr}
Suppose that, in practice, we have $B$ independent tasks corresponding to unobservable $f_1,\ldots,f_B$ that are i.i.d. drawn from $F$. 
For task $b$, suppose we have $n_b$ context data $D_{n_b}^b=(Z_1^b,\ldots,Z_{n_b}^b)$,
and $N_b$ test data $Z_{n_b+1}^b,\ldots,Z_{n_b+N_b}^b$, where $Z_i^b=(X_i^b,Y_i^b)$ with $X_1^b,\ldots,X_{n_b+N_b}^b\overset{i.i.d.}{\sim}\mu$ and
\begin{equation}\label{eq:model:S+N}
Y_i^b=f_b(X_i^b)+\varepsilon_i^b,\,\,\,\,i=1,\ldots,n_b+N_b.
\end{equation}
The errors $\varepsilon_i^b$'s are i.i.d. with mean zero and finite variance $\sigma_\varepsilon^2$, independent of $f_b,X_1^b,\ldots,X_{n_b+N_b}^b$. 
Let $\widehat{\pi}_{n_b}^b=n_b^{-1}\sum_{i=1}^{n_b}\delta_{Z_i^b}$ be the empirical measure based on $D_{n_b}^b$. Similar to (\ref{alt:expression:M_*}), it can be shown that 
\[
\E[Y_{n_b+i}^b|\widehat{\pi}^b_{n_b},X_{n_b+i}^b=x]=M_*(\widehat{\pi}_{n_b}^b,x).
\]
That is, $M_*$ is also the conditional mean of the test response $Y_{n+i}^b$ given $\widehat{\pi}_{n_b}^b,X_{n+i}^b$. So we may estimate $M_*$ by minimizing the sum of squared residuals based on the test data.
Specifically, we propose the following measure-indexed kernel ridge regression (MI-KRR):
\begin{eqnarray}\label{eqn:hyper:krr}
\widehat{M}_{\textbf{n},\textbf{N},B,\lambda}&=&\argmin_{M\in\cM}\ell_{\textbf{n},\textbf{N},B,\lambda}(M)\nonumber\\
&\equiv&\argmin_{M\in\cM}\left\{\frac{1}{2N}\sum_{b=1}^B\sum_{i=1}^{N_b}|Y_{n_b+i}^b-M(\widehat{\pi}_{n_b}^b,X_{n_b+i}^b)|^2+\frac{\lambda}{2}\|M\|_\cM^2\right\},
\end{eqnarray}
where $\lambda>0$ is the smoothing parameter and $\|\cdot\|_\cM$ is the RKHS norm of $\cM$, 
$\textbf{n}=(n_1,\ldots,n_B)$ and $\textbf{N}=(N_1,\ldots,N_B)$ represent the sizes of context data sets and test data sets, respectively, and $N=N_1+\cdots+N_B$ is the total number of test data points. 
Given a new task involving an in-context sample encoded by an empirical measure $\widehat{\pi}$ and a query input $x$, we can directly predict the new response by $\widehat{M}_{\textbf{n},\textbf{N},B,\lambda}(\widehat{\pi},x)$. Thus, the proposed MI-KRR estimator can automatically adapt to new task without refitting or retraining. 

By representer theorem \citep{Wahba:90}, $\widehat{M}_{\textbf{n},\textbf{N},B,\lambda}$ has the form
\[
\widehat{M}_{\textbf{n},\textbf{N},B,\lambda}(\widehat{\pi}, x) = \sum_{b=1}^{B}\sum_{i=1}^{N_b} c_i^b \, R((\widehat{\pi}, x),(\widehat{\pi}_{n_b}^b, X_{n_b+i}^b)),
\]
where the coefficient vector $\textbf{c} = (c_1^1,\ldots,c_{N_1}^1,\ldots,c_1^B,\ldots,c_{N_B}^B)^\top$ is given by 
\[
\textbf{c} = (\textbf{R}_{N\times N}+ N\lambda \textbf{I}_{N\times N})^{-1}\textbf{Y},
\]
where $\textbf{R}_{N\times N}$ is the $N\times N$ Gram matrix associated with the reproducing kernel $R$ of the RKHS $\cM$, i.e., the $((b,i),(b',i'))$-th element of $\textbf{R}_{N\times N}$ is $R((\widehat{\pi}_{n_b}^b,X_{n_b+i}^b),(\widehat{\pi}_{n_{b'}}^{b'},X_{n_{b'}+i'}^{b'}))$, and $\textbf{Y}$ is the response vector defined by $\textbf{Y} = (Y_{n_1+1}^1,\ldots,Y_{n_1+N_1},\ldots,Y_{n_B+1}^B,\ldots,Y_{n_B+N_B}^B)^\top$.
Motivated by classic smoothing spline regression \citep{gu2013}, we suggest to select $\lambda$ using generalized cross-validation (GCV). 
Define the smoothing matrix
\[
\mathbf{H}_\lambda=\mathbf{R}_{N\times N}(\mathbf{R}_{N\times N}+N\lambda\mathbf{I}_{N\times N})^{-1}.
\]
The GCV criterion is
\[
\operatorname{GCV}(\lambda)
=\frac{N^{-1}|(\mathbf{I}_{N\times N}-\mathbf{H}_\lambda)\mathbf{Y}|^2}
{{1-N^{-1}\operatorname{trace}(\mathbf{H}_\lambda)}^2},\,\,\,\,\lambda>0.
\]
We then choose
$\widehat{\lambda}_{\mathrm{GCV}}
=\arg\min_{\lambda}\operatorname{GCV}(\lambda)$.

\section{Asymptotic Properties}
In this section, we will derive the convergence rate and asymptotic distribution for $\widehat{M}_{\textbf{n},\textbf{N},B,\lambda}$.
Before stating our main results, let us make some technical preparations.

\subsection{Spectral Property of MI-RKHS}
In this section, we will establish the spectral property of $\cM$ which is crucial for asymptotic analysis.
We make the following technical assumption.
\begin{assumption}\label{A2}
$C_\Pi\triangleq\sup_{\pi\in\mcP(\mcZ)}
\left\|R^{\langle\Pi\rangle}(\pi,\pi)\right\|_{\textrm{op}}<\infty$, where $\|\cdot\|_{\textrm{op}}$ is the operator norm.
\end{assumption}
Assumption \ref{A2} bounds the operator norm of the kernel $R^{\langle\Pi\rangle}$ evaluated at arbitrary measures in $\mcP(\mcZ)$. Intuitively, it controls how much the measure-indexed kernel can amplify differences between context distributions, ensuring that the proposed MI-KRR estimator remains stable across different in-context samples. It is the measure-valued analogue of a standard bounded-kernel assumption in classical RKHS literature.
For $M,M'\in\cM$, define a quadratic form
\[
V(M,M')=\frac{1}{N}\sum_{b=1}^BN_b\E[M(\widehat{\pi}_{n_b}^b,X_{n_b+1}^b)M'(\widehat{\pi}_{n_b}^b,X_{n_b+1}^b)].
\]
By (\ref{reproducing:property:R}), we have the following
\begin{eqnarray*}
V(M,M')&=&\frac{1}{N}\sum_{b=1}^BN_b\E\left[\langle M, R_{(\widehat{\pi}_{n_b}^b,X_{n_b+1}^b)}\rangle_\cM\langle M', R_{(\widehat{\pi}_{n_b}^b,X_{n_b+1}^b)}\rangle_\cM\right]\\
&=&\left\langle\frac{1}{N}\sum_{b=1}^BN_b\E\left[R_{(\widehat{\pi}_{n_b}^b,X_{n_b+1}^b)}\otimes R_{(\widehat{\pi}_{n_b}^b,X_{n_b+1}^b)}\right]M,M'\right\rangle_\cM\equiv\langle\mcC M,M'\rangle_\cM,
\end{eqnarray*}
where 
\[
\mcC_{\textbf{n},\textbf{N},B}\triangleq \frac{1}{N}\sum_{b=1}^BN_b\E\left[R_{(\widehat{\pi}_{n_b}^b,X_{n_b+1}^b)}\otimes R_{(\widehat{\pi}_{n_b}^b,X_{n_b+1}^b)}\right]
\]
is a nonnegative definite rank-one operator from $\cM$ to $\cM$. 
To ease notation, we simply denote $\mcC=\mcC_{\textbf{n},\textbf{N},B}$.
By Assumption \ref{A2}, (\ref{reproducing:kernel:M}) and (\ref{reproducing:property:R}), 
for any $(\widehat{\pi},x)\in\mcP(\mcZ)\times\cX$,
\begin{eqnarray}\label{R:norm:ineqality}
\|R_{(\widehat{\pi},x)}\|_\cM^2&=&
R((\widehat{\pi},x),(\widehat{\pi},x))\,\,\,\,\,\,\,\, \textrm{(by (\ref{reproducing:property:R}))}\nonumber\\
&=&\langle R^{\langle\Pi\rangle}_{\widehat{\pi}} R^{\langle X\rangle}_x, R^{\langle\Pi\rangle}_{\widehat{\pi}} R^{\langle X\rangle}_x\rangle_{\cH_\Pi}\,\,\,\,\,\,\,\, \textrm{(by (\ref{reproducing:kernel:M}))}\nonumber\\
&=&\langle
R^{\langle X\rangle}_x,
R^{\langle\Pi\rangle}(\widehat{\pi},\widehat{\pi})
R^{\langle X\rangle}_x\rangle_{\cH_X}\,\,\,\,\,\,\,\, \textrm{(by Property~\ref{eqn:reproducing:property:ii})}\nonumber\\
&\leq&
\|R^{\langle\Pi\rangle}(\widehat{\pi},\widehat{\pi})\|_{\textrm{op}}
\|R^{\langle X\rangle}_x\|_{\cH_X}^2\nonumber\\
&=&\|R^{\langle\Pi\rangle}(\widehat{\pi},\widehat{\pi})\|_{\textrm{op}}R^{\langle X\rangle}(x,x)
\le C_\Pi C_X<\infty\,\,\,\,\,\,\,\, \textrm{(by Assumptions \ref{A0} and \ref{A2}).}    
\end{eqnarray}
Consequently,
\[
\textrm{trace}(\mcC)=\frac{1}{N}\sum_{b=1}^BN_b\E\left[\|R_{(\widehat{\pi}_{n_b}^b,X_{n_b+1}^b)}\|_\cM^2\right]\le C_\Pi C_X,
\]
i.e., $\mcC$ is a nonnegative trace-class operator. 
Let $\textrm{ker}(\mcC)=\{M\in\cM: \mcC M=0\}$ be the kernel space of $\mcC$,
and $\textrm{ker}(\mcC)^\perp$ its orthogonal complement with respect to inner product $\langle\cdot,\cdot\rangle_\cM$.
The range of $\widehat{\pi}_{n_b}^b$'s may not be the entire range $\mcP(\mcZ)$, so it is possible that $\mcC M=0$ for some nonzero $M$,
i.e., $\textrm{ker}(\mcC)\neq\{0\}$.
By spectral theorem in Hilbert spaces (see Corollary 5.4 of \cite{conway1990course}), there exist a positive nondecreasing sequence
$\{\rho_k\}_{k\ge1}$ and a basis $\{\varphi_k\}_{k\ge1}\subseteq\textrm{ker}(\mcC)^\perp$ such that
\begin{equation}\label{simultaneous:diag}
V(\varphi_k,\varphi_l)=\langle\mcC\varphi_k,\varphi_l\rangle_\cM=\delta_{kl},\,\,\langle\varphi_k,\varphi_l\rangle_\cM=\rho_k\delta_{kl},\,\,k,l\ge 1,
\end{equation}
where $\delta_{kl}$ are Kronecker's notation. 
When $\textrm{ker}(\mcC)=\{0\}$ so that $\textrm{ker}(\mcC)^\perp=\cM$, $\{\varphi_k\}_{k\ge1}$ becomes the basis for $\cM$. Our results apply to general $\textrm{ker}(\mcC)$, either null or non-null.
For technical convenience, we further assume that these basis functions are uniformly bounded.
\begin{assumption}\label{A3} 
There exists a constant $C_\varphi$ which is free of $\textbf{n},\textbf{N},B$ such that $\sup_{k\ge1}\|\varphi_k\|_\infty\le C_\varphi$. 
\end{assumption}
Assumption \ref{A3} requires that the basis functions $\varphi_k$ used to expand the RKHS $\cM$ are uniformly bounded in sup-norm. It is a standard regularity condition in nonparametric literature that facilitates the establishment of certain maximal inequalities. The requirement that $C_\varphi$ be free of $\mathbf{n}, \mathbf{N}, B$ ensures that the bound does not deteriorate as these indices grow.

For $M,M'\in\cM$, define $\langle M,M'\rangle=V(M,M')+\lambda\langle M,M'\rangle_\cM$.
Then $\langle\cdot,\cdot\rangle$ and $\langle\cdot,\cdot\rangle_\cM$ are equivalent inner products for $\cM$, hence,
$\cM$ under $\langle\cdot,\cdot\rangle$ is also an RKHS.
To see this, for any $M\in\cM$, consider the orthogonal decomposition $M=M_1+M_2$ with $M_1\in\textrm{ker}(\mcC)^\perp$ and $M_2\in\textrm{ker}(\mcC)$. Write $M_1=\sum_{k\ge1}m_k\varphi_k$. Then
\[
 V(M_1,M_2)=\sum_{k\ge1}m_k^2\le\rho_1^{-1}\sum_{k\ge1}\rho_k m_k^2=\rho_1^{-1}\|M_1\|_\cM^2.   
\]
Since $\|M\|_\cM^2=\|M_1\|_\cM^2+\|M_2\|_\cM^2\ge\|M_1\|_\cM^2$, we have
\begin{eqnarray*}
\|M\|^2&\triangleq&V(M,M)+\lambda\|M\|_\cM^2=V(M_1,M_1)+\lambda\|M\|_\cM^2\\
&\le& \rho_1^{-1}\|M_1\|_\cM^2+\lambda\|M\|_\cM^2\le (\rho_1^{-1}+\lambda)\|M\|_\cM^2.    
\end{eqnarray*}
On the other hand, $\|M\|^2\ge\lambda\|M\|_\cM^2$, we therefore get the equivalence of $\|\cdot\|$ and $\|\cdot\|_\cM$.

Let $K:(\mcP(\mcZ)\times\cX)^2\to\mbbR$ be the reproducing kernel for $\cM$ under $\langle\cdot,\cdot\rangle$, which satisfies the following reproducing property: for any $(\widehat{\pi},x)\in\mcP(\mcZ)\times\cX$ and $M\in\cM$,
\[
K_{(\widehat{\pi},x)}\triangleq K((\cdot,\cdot),(\widehat{\pi},x))\in\cM,\,\,\,\,\langle M, K_{(\widehat{\pi},x)}\rangle=M(\widehat{\pi},x).
\]
For any $M\in\cM$, $\lambda\langle M,\cdot\rangle_\cM$ is a bounded linear functional on $(\cM,\|\cdot\|)$. By Riesz representation theorem,
there is a bounded linear operator $P_\lambda: \cM\to\cM$ such that, for any $M,M'\in\cM$,
\[
\langle P_\lambda M,M'\rangle=\lambda\langle M,M'\rangle_\cM.
\]

\subsection{Convergence Rate and Asymptotic Normality}\label{sec:asymptotics}
In this section, we derive the convergence rate and asymptotic distribution for $\widehat{M}_{\textbf{n},\textbf{N},B,\lambda}$.
For $b=1,\ldots,B$, let 
\[
V_b^2(n_b)=\E[|f_b(X_{n_b+1}^b)-\E[f_b|\widehat{\pi}_{n_b}^b](X_{n_b+1}^b)|^2],
\]
and $V^2(\textbf{n})=\max_{1\le b\le B}V_b^2(n_b)$. Specifically, $V_b^2(n_b)$ measures the posterior uncertainty of $f_b$
given the $b$-th context data, and $V^2(\textbf{n})$ represents the largest such uncertainty across all $B$ tasks.
Let $e_{n_b,i}^b=Y_{n_b+i}^b-M_*(\widehat{\pi}_{n_b}^b,X_{n_b+i}^b), i=1,\ldots,N_b$ denote the prediction errors, 
which measure the discrepancies between test responses and their predictions based on the context data.
Recall also that $\varepsilon_i^b$'s denote the observation errors defined in (\ref{eq:model:S+N}).
We further assume that these error terms have finite second moments which is a mild condition commonly used in nonparametric literature.
\begin{assumption}\label{A4}
There are positive constants $\sigma_e^2$ and $\sigma_\varepsilon^2$ such that, for $b=1,\ldots,B$,
\[
\textrm{$\E\left[|e_{n_b,i}^b|^2\big|\widehat{\pi}_{n_b}^b,X_{n_b+i}^b\right]\le \sigma_e^2$ almost surely, and $\E[|\varepsilon_i^b|^2]=\sigma_\varepsilon^2$.}    
\]
\end{assumption}
Let 
\[
\textrm{
$\cM_h=\{M\in\cM: \|M\|\le C_\varphi^{-1}h^{1/2}\}$
and 
$a_h=\int_0^1\sqrt{\log(1+D(\cM_h,\varepsilon,\|\cdot\|_{\infty}))}d\varepsilon$,}
\]
where $h=\left(\sum_{k\ge1}\frac{1}{1+\lambda\rho_k}\right)^{-1}$
and $D(\cM_h,\varepsilon,\|\cdot\|_{\infty})$ is the $\varepsilon$-covering number of $\cM_h$ with respect to supremum norm $\|\cdot\|_\infty$. 
The following theorem provides a convergence rate for $\widehat{M}_{\textbf{n},\textbf{N},B,\lambda}$.
\begin{theorem}\label{thm:rate:loo}
Suppose that Assumptions \ref{A0}-\ref{A4} are satisfied and as $\textbf{n},\textbf{N},B\to\infty$ (meaning that all $n_b$'s, $N_b$'s, $B$ tend to infinity), 
\begin{equation}\label{eq:rate:condition}
a_h\sqrt{\sum_{b=1}^B N_b^2}=o(Nh).
\end{equation}
Then 
$\|\widehat{M}_{\textbf{n},\textbf{N},B,\lambda}-M_*\|=O_\mbbP\left(\lambda^{1/2}+U(\textbf{n},\textbf{N},B,\lambda)(Nh)^{-1/2}\right)$,
where 
\[
U(\textbf{n},\textbf{N},B,\lambda)=\max\{1,(\lambda\max_{1\le b\le B} N_b)^{1/2},V(\textbf{n})(\max_{1\le b\le B}N_b)^{1/2}\}.
\]
\end{theorem}
We will demonstrate the rate of convergence of Theorem \ref{thm:rate:loo} under a concrete setting.
For technical convenience, consider the balanced case $n_b=n$ and $N_b=T$ for all $b=1,\ldots,B$. 
That is, each task has the
same context size and the same number of test observations. The total number of test data is therefore $N=TB$.
Suppose $\rho_k\asymp k^{2\gamma}$, i.e.,
there exist constants $c_1,c_2>0$ and $\gamma>1/2$ such that
\begin{equation}\label{eigen:decay}
c_1k^{2\gamma}\le\rho_k\le c_2k^{2\gamma},\,\,\,\, k\ge1.
\end{equation}
Under (\ref{eigen:decay}), it can be checked that $\lambda\asymp h^{2\gamma}$,
and the eigenvalues of
$\mcC$ satisfy an order $k^{-2\gamma}$.
Suppose 
\begin{equation}\label{covering:number:order}
\textrm{$\log{D(\mathbb{B}_\cM,\varepsilon,\|\cdot\|_\infty)}\lesssim\varepsilon^{-1/\gamma}$,\,\,for $\varepsilon>0$,}
\end{equation}
where $\mathbb{B}_\cM=\{M\in\cM: \|M\|_\cM\le1\}$ is the unit ball of $\cM$.
The exponent $1/\gamma$ in (\ref{covering:number:order}) is consistent with
the eigenvalue decay of $\mcC$, which reflects a well-known relationship
between the metric entropy of an RKHS unit ball
and its kernel eigenvalue decay \citep{carl1990entropy}.
Since $\cM_h\subset C_\varphi^{-1}(h/\lambda)^{1/2}\mathbb{B}_\cM$,
we have 
\begin{eqnarray*}
\log{D(\cM_h,\varepsilon,\|\cdot\|_\infty)}&\le&\log{D(C_\varphi^{-1}(h/\lambda)^{1/2}\mathbb{B}_\cM,\varepsilon,\|\cdot\|_\infty)}\\
&=&\log{D(\mathbb{B}_\cM,C_\varphi(h/\lambda)^{-1/2}\varepsilon,\|\cdot\|_\infty)}\lesssim (h/\lambda)^{1/(2\gamma)}\varepsilon^{-1/\gamma}.
\end{eqnarray*}
Consequently, $a_h=O(h^{(1-2\gamma)/(4\gamma)})$. 
When $h\asymp N^{-1/(2\gamma+1)}$ and $T=o(B^{(2\gamma-1)^2/(6\gamma-1)})$ (i.e., the number of tasks $B$ should not be too small compared with test data size $T$), 
condition (\ref{eq:rate:condition}) is satisfied. 
Meanwhile, it can be checked that 
\[
\lambda\max_{1\le b\le B}N_b=\lambda T\asymp TN^{-2\gamma/(2\gamma+1)}=(TB^{-2\gamma})^{1/(2\gamma+1)}=O(1).
\]
If we further assume that $V(\textbf{n})=O(T^{-1/2})$, i.e., the posterior uncertainty of $f_b$ is uniformly small compared with $T$,
then $U(\textbf{n},\textbf{N},B,\lambda)=O(1)$. Therefore, Theorem \ref{thm:rate:loo} implies that
\begin{equation}\label{optimal:rate:hat:M}
\|\widehat{M}_{\textbf{n},\textbf{N},B,\lambda}-M_*\|=O_\mbbP\left(\lambda^{1/2}+(Nh)^{-1/2}\right)=O_\mbbP\left(N^{-\gamma/(2\gamma+1)}\right).
\end{equation}
The rate is minimax optimal since there are $N=TB$ test data involved in the MI-KRR (\ref{eqn:hyper:krr}).
In the classic distributed learning \citep{zhang2013jmlr,shang2017jmlr,shang:jcgs:2019,shang2019jmlr,banerjee2019aos,shang:ejs:2020,szabo2020aos}, all task-specific regression functions coincide, i.e., $f_1=\cdots=f_B=f_*$, where $f_*$ is unknown deterministic. Consequently, $f_b$'s are sampled from the point mass measure $F=\delta_{f_*}$ which leads to $V(\textbf{n})=0$ (regardless of $\textbf{n}$). We can therefore choose $n\asymp T$ in each task so that the context and test sample sizes are of the same order.
Since $N=TB$, the total number of observations across all tasks, including both context and test data, is of order $2N$. Thus, the rate in (\ref{optimal:rate:hat:M}) is minimax optimal even when accounting for the context data. More generally, $V(\mathbf{n})\neq0$, but the following Lemma \ref{lem:hat:V_n} establishes a convergence rate for it. This allows the framework to accommodate heterogeneous tasks whose underlying regression functions $f_b$ are similar but not necessarily identical, connecting our results to meta-learning and transfer learning \citep{li2021jrssb,li2023jasa,li2024jasa,ye2023jasa,cai2021aos,cai2024arxiv,shang2025bootstrapnonparametricinferencedata}.

Suppose that the reproducing kernel $R^{\langle X\rangle}$ satisfies a Mercer decomposition
\begin{equation}\label{mercer:cH_X}
R^{\langle X\rangle}(x,x')=\sum_{k\ge1}\varphi^{\langle X\rangle}_k(x)\varphi^{\langle X\rangle}_k(x')/\rho^{\langle X\rangle}_k,\,\,\,x,x'\in\cX,
\end{equation}
where $\varphi^{\langle X\rangle}_k$ are orthonormal basis for $L^2(\mu)$
with $C_{\varphi^{\langle X\rangle}}=\sup_{k\ge1}\|\varphi^{\langle X\rangle}\|_\infty<\infty$ and 
$\rho^{\langle X\rangle}_k\asymp k^{2\beta}$ is a positive sequence with $\beta>1/2$.
The following lemma establishes a nonasymptotic rate for $V_b(n_b)$.
\begin{lemma}\label{lem:hat:V_n}
Suppose that $\varepsilon_i^b$'s have finite fourth moments, 
Assumption \ref{A1} holds and $R^{\langle X\rangle}$ has decomposition (\ref{mercer:cH_X}), then
 \[
V_b(n_b)^2\le C_1(1+\E[\|f_b\|_{\cH_X}^2])n^{-\frac{2\beta}{2\beta+1}},\,\,\,\,b=1,\ldots,B, 
\]
where $C_1$ is a constant.
\end{lemma}
By Lemma~\ref{lem:hat:V_n}, $V(\mathbf{n}) = O(n^{-\beta/(2\beta+1)})$, so the condition $V(\mathbf{n})\,T^{1/2} = O(1)$ is satisfied whenever $T = O(n^{2\beta/(2\beta+1)})$. As $\beta \to \infty$, the exponent $2\beta/(2\beta+1) \to 1$, permitting $T$ as large as $O(n)$ and hence a total sample size of order $2N$ across all tasks. Thus the rate in \eqref{optimal:rate:hat:M} remains nearly minimax optimal even after accounting for the context observations.

In the end of this section, we establish
the asymptotic normality for
$\widehat{M}_{\mathbf{n},\mathbf{N},B,\lambda}$.
Let $\widehat{\pi}\in\mcP(\mcZ)$ be a
nonrandom probability measure, possibly
a realization of an empirical measure,
and $x\in\cX$ be a nonrandom vector.
We regard $(\widehat{\pi},x)$ as the prompt
of a new task, representing the context
distribution and a query input, respectively.
The pointwise prediction of $M_*$ at
$(\widehat{\pi},x)$ is
$M_*(\widehat{\pi},x)$, which can be estimated by
$\widehat{M}_{\mathbf{n},\mathbf{N},B,\lambda}
(\widehat{\pi},x)$. The following theorem establishes its asymptotic normality. 

\begin{theorem}\label{thm:asymp:normality}
Suppose that Assumptions \ref{A0}-\ref{A4} are satisfied and, as $\textbf{n},\textbf{N},B\to\infty$, (\ref{eq:rate:condition}) holds and
\begin{eqnarray}
&&U(\textbf{n},\textbf{N},B,\lambda)=O(1),\label{thm:asymp:normality:limit:condition:1}\\
&&\frac{1}{\sigma_{(\widehat{\pi},x)}^2}\sum_{k=1}^\infty\frac{h|\varphi_k(\widehat{\pi},x)|^2}{(1+\lambda\rho_k)^2}\to1,\label{thm:asymp:normality:limit:condition:2}\\
&&\sqrt{Nh}(P_\lambda M_*)(\widehat{\pi},x)=o(1),\label{thm:asymp:normality:limit:condition:3}\\
&&\frac{a_h\sqrt{\sum_{b=1}^BN_b^2}}{\sqrt{N}h}\left(\lambda^{1/2}+(Nh)^{-1}\right)=o(1),\label{thm:asymp:normality:limit:condition:4}\\
&&\sum_{b=1}^BN_b^2V(\textbf{n})^2=o(Nh), \label{thm:asymp:normality:limit:condition:5}
\end{eqnarray}
where $\sigma_{\widehat{\pi},x}^2$ is a constant possibly depending on $\widehat{\pi},x$. Then 
\[
\frac{\sqrt{Nh}}{\sigma_\varepsilon\sigma_{(\widehat{\pi},x)}}(\widehat{M}_{\textbf{n},\textbf{N},B,\lambda}(\widehat{\pi},x)-M_*(\widehat{\pi},x))\overset{d}{\to}N(0,1).
\]
\end{theorem}
Theorem~\ref{thm:asymp:normality} establishes a pointwise CLT for $\widehat{M}_{\mathbf{n},\mathbf{N},B,\lambda}(\widehat{\pi},x)$, providing the asymptotic foundation for inference on $M_*(\widehat{\pi},x)$. Notably, the estimator adapts to a new task $\widehat{\pi}$ without any refitting of the model, yet still achieves asymptotic normality.
Condition (\ref{thm:asymp:normality:limit:condition:2}) says that the asymptotic variance of $\widehat{M}_{\textbf{n},\textbf{N},B,\lambda}(\widehat{\pi},x)$ converges to $\sigma_{\widehat{\pi},x}^2$, which is a common assumption in establishing CLT-type results.
Condition (\ref{thm:asymp:normality:limit:condition:3}) is known as the undersmoothing condition, which is useful to remove the asymptotic bias in $\widehat{M}_{\textbf{n},\textbf{N},B,\lambda}(\widehat{\pi},x)$. To be more explicit, suppose $M_*=\sum_{k=1}^\infty m_k\varphi_k$,
then 
\[
P_\lambda M_*=\sum_{k=1}^\infty m_k P_\lambda\varphi_k=\sum_{k=1}^\infty \frac{\lambda\rho_km_k}{1+\lambda\rho_k}\varphi_k.
\]
If $\sum_{k=1}^\infty m_k^2\rho_k^{1+\epsilon}<\infty$ for some constant $\epsilon\in(1/(2\gamma),1)$,
i.e., $M_*$ is slightly smoother than the regularity of $\cH_X$, then
\[
\|P_\lambda M_*\|^2=\sum_{k=1}^\infty m_k^2\frac{(\lambda\rho_k)^2}{1+\lambda\rho_k}=\sum_{k=1}^\infty m_k^2(\lambda\rho_k)^{1+\epsilon}\frac{(\lambda\rho_k)^{1-\epsilon}}{1+\lambda\rho_k}=O(\lambda^{1+\epsilon}).
\]
When $h\asymp N^{-1/(2\gamma+1)}$, by Lemma \ref{boundedness:lemma},
\[
\sqrt{Nh}|(P_\lambda M_*)(\widehat{\pi},x)|\le \sqrt{Nh}C_\varepsilon h^{-1/2}\|P_\lambda M_*\|=O(\sqrt{N}N^{-\gamma(1+\epsilon)/(2\gamma+1)})=o(1),
\]
which verifies Condition (\ref{thm:asymp:normality:limit:condition:3}).
It can be directly checked that Condition (\ref{thm:asymp:normality:limit:condition:4}) holds when $\gamma>(3+\sqrt{5})/4$ and 
$T=o(B^{(4\gamma^2-6\gamma+1)/(8\gamma-1)})$; the larger value of $\gamma$ implies a smoother RKHS $\cM$.
By Lemma \ref{lem:hat:V_n}, it can be checked that (\ref{thm:asymp:normality:limit:condition:4}) holds if
$T^{2(\gamma+1)}B=o(n^{2\beta(2\gamma+1)/(2\beta+1)})$. 

In practice, we suggest to estimate $\sigma_{(\widehat{\pi},x)}^2$ by
\[
\widehat{\sigma}_{(\widehat{\pi},x)}^2=
\sum_{k=1}^\infty\frac{h|\varphi_k(\widehat{\pi},x)|^2}{(1+\lambda\rho_k)^2}.
\]
Suppose $D_n^{\textrm{new}}=\{Z_1^{\textrm{new}},\ldots,Z_n^{\textrm{new}}\}$ is an in-context sample from a new task
with $Z_i^{\textrm{new}}=(X_i^{\textrm{new}},Y_i^{\textrm{new}})$ generated from the following model
\[
Y_i^{\textrm{new}}=f_{\textrm{new}}(X_i^{\textrm{new}})+\varepsilon_i^{\textrm{new}},\,\,i=1,\ldots,n.
\]
Let $\widehat{\pi}^{\textrm{new}}_n$ be the empirical distribution of $D_n^{\textrm{new}}$.
Therefore, $M_*(\widehat{\pi},x)=\E[Y^{\textrm{new}}|\widehat{\pi}_n^{\textrm{new}}=\widehat{\pi},X^{\textrm{new}}=x]$.
Based on Theorem \ref{thm:asymp:normality}, we suggest the following $100\%(1-\alpha)$ confidence interval for $M_*(\widehat{\pi},x)$:
\[
\widehat{M}_{\mathbf{n},\mathbf{N},B,\lambda}
(\widehat{\pi},x)\pm z_{1-\alpha/2}\frac{\widehat{\sigma}_{(\widehat{\pi},x)}\widehat{\sigma}_\varepsilon}{\sqrt{Nh}},
\]
where $z_{1-\alpha/2}$ is the $(1-\alpha/2)$-th quantile of the standard normal distribution, and
\[
\widehat{\sigma}_\varepsilon^2 = N^{-1}\sum_{b=1}^B\sum_{i=1}^{N_b}
(Y_{n_b+i}^b - \widehat{M}_{\mathbf{n},\mathbf{N},B,\lambda}
(\widehat{\pi}_{n_b}^b, X_{n_b+i}^b))^2
\]
is a consistent estimator of $\sigma_\varepsilon^2$.
Under $\widehat{\pi}_n^{\textrm{new}}=\widehat{\pi}$ and $X^{\textrm{new}}=x$, we have
\begin{eqnarray*}
Y^{\textrm{new}}-\widehat{M}_{\mathbf{n},\mathbf{N},B,\lambda}(\widehat{\pi},x)&=&f_{\textrm{new}}(x)-\widehat{M}_{\mathbf{n},\mathbf{N},B,\lambda}(\widehat{\pi},x)+\varepsilon^{\textrm{new}}\\
&=&f_{\textbf{new}}(x)-M_*(\widehat{\pi},x)+M_*(\widehat{\pi},x)-\widehat{M}_{\mathbf{n},\mathbf{N},B,\lambda}(\widehat{\pi},x)+\varepsilon^{\textrm{new}}\\
&=&O_\mbbP\left(n^{-\beta/(2\beta+1)}\right)+M_*(\widehat{\pi},x)-\widehat{M}_{\mathbf{n},\mathbf{N},B,\lambda}(\widehat{\pi},x)+\varepsilon^{\textrm{new}},
\end{eqnarray*}
where the $O_\mbbP\left(n^{-\beta/(2\beta+1)}\right)$ term in the last equation follows by Lemma \ref{lem:hat:V_n}.
Note that $\varepsilon^{\textrm{new}}$ is independent of $\widehat{M}_{\mathbf{n},\mathbf{N},B,\lambda}(\widehat{\pi},x)$.
If further $\varepsilon^{\textrm{new}}\sim N(0,\sigma_\varepsilon^2)$, by Theorem \ref{thm:asymp:normality},  
\[
M_*(\widehat{\pi},x)-\widehat{M}_{\mathbf{n},\mathbf{N},B,\lambda}(\widehat{\pi},x)+\varepsilon^{\textrm{new}}\overset{a.d.}{\sim}N\left(0,\sigma_\varepsilon^2\sigma_{(\widehat{\pi},x)}^2/(Nh)+\sigma_\varepsilon^2\right),
\]
where ``$\overset{a.d.}{\sim}$'' means ``approximately distributed.''
If $n^{-\beta/(2\beta+1)}=o((Nh)^{-1/2})$, the bias term is asymptotically negligible.
Then we drop the $O_\mbbP\left(n^{-\beta/(2\beta+1)}\right)$ term, so that
\[
Y^{\textrm{new}}-\widehat{M}_{\mathbf{n},\mathbf{N},B,\lambda}(\widehat{\pi},x)\overset{a.d.}{\sim}N\left(0,\sigma_\varepsilon^2\sigma_{(\widehat{\pi},x)}^2/(Nh)+\sigma_\varepsilon^2\right).
\]
Therefore, we suggest the following $100\%(1-\alpha)$ prediction interval for
the new response $Y_{\mathrm{new}}$:
\[
\widehat{M}_{\mathbf{n},\mathbf{N},B,\lambda}(\widehat{\pi},x)
\pm z_{1-\alpha/2}
\sqrt{\frac{\widehat{\sigma}_{(\widehat{\pi},x)}^2\widehat{\sigma}_\varepsilon^2}{Nh}+\widehat{\sigma}_\varepsilon^2}.
\]

\subsection{Example: Latent Gaussian Process Model}\label{sec:GP}
In this section, we consider a latent Gaussian process model which will validate the proposed MI-RKHS framework.
Specifically, we will describe an explicit construction of the RKHS $\cH_\Pi$ as well as the corresponding reproducing kernel $R^{\langle\Pi\rangle}$ to which $H_*$ belongs. 
Recall that $R^{\langle\Pi\rangle}$ and $R^{\langle X\rangle}$ jointly determine the reproducing kernel $R$ for the MI-RKHS via (\ref{reproducing:kernel:M}),
hence, such explicit construction is useful for the application of the proposed MI-KRR (\ref{eqn:hyper:krr}).
Suppose that $\cH_X$ is closed under pointwise multiplication, i.e., there exists a constant $C_d$ such that for any $f_1,f_2\in\cH_X$,
\begin{equation}\label{eq:multiplication:preserving}
\|f_1f_2\|_{\cH_X}\le C_d\|f_1\|_{\cH_X}\|f_2\|_{\cH_X}.
\end{equation}
This holds, for example, when $\cH_X$ is a Sobolev space of order greater than $d/2$ on a compact domain $\cX\subset\mbbR^d$ \citep{adams2003sobolev}.
Suppose that the latent random function $f$ satisfies $f=f_*+W$, where $f_*\in\cH_X$ is unknown deterministic mean function
and $W$ is a centered Gaussian process with unknown deterministic covariance function
$C_*$ such that $\E[\|W\|_{\cH_X}^2]<\infty$, which implies $f\in\cH_X$ almost surely.
Moreover, for any $x\in\cX$, by the reproducing property,
\begin{eqnarray}
\|C_*(\cdot,x)\|_{\cH_X}&=&\|\E[W(\cdot)W(x)]\|_{\cH_X}\le \E[|W(x)|\cdot\|W\|_{\cH_X}]\nonumber\\
&\le&\E[\|W\|_{\cH_X}^2\|R^{\langle X\rangle}_x\|_{\cH_X}]\le  C_X^{1/2}\E[\|W\|_{\cH_X}^2]<\infty,\label{C*:H_X}    
\end{eqnarray}
therefore, $C_*(\cdot,x)$ belongs to $\cH_X$.
Let $\Gamma_*=\E[W\otimes W]$. Since $\textrm{trace}(\Gamma_*)=\E[\|W\|_{\cH_X}^2]<\infty$, $\Gamma_*$ is trace-class, hence bounded from $\cH_X$ to $\cH_X$. We also have
\begin{equation}\label{eqn:C_*:Gamma_*}
C_*(\cdot,x)=\E[W(\cdot)W(x)]=\E[\langle W,R^{\langle X\rangle}_\cdot\rangle_{\cH_X}\langle W,R^{\langle X\rangle}_x\rangle_{\cH_X}]=\langle\Gamma_* R^{\langle X\rangle}_x,R^{\langle X\rangle}_\cdot\rangle_{\cH_X}=(\Gamma_* R^{\langle X\rangle}_x)(\cdot).
\end{equation}
For $\pi\in\mcP(\mcZ)$, define
\begin{equation}\label{eqn:G:hat}
G_\pi=\int_\mcZ(y-f_*(x))\Gamma_*^{1/2}R^{\langle X\rangle}_xd\pi(x,y),\,\,\,\,
J_\pi=\int_\cX (\Gamma_*^{1/2}R^{\langle X\rangle}_x)\otimes(\Gamma_*^{1/2}R^{\langle X\rangle}_x)d\pi^X(x),\,\,g\in\cH_X,
\end{equation}
where $\pi^X$ is the marginal of $X$ under $\pi$.
By Cauchy-Schwartz inequality and Assumption \ref{A0},
\begin{eqnarray*}
\|G_\pi\|_{\cH_X}^2&\le&\int_\mcZ|y-f_*(x)|^2d\pi(x,y)\|\Gamma_*^{1/2}R^{\langle X\rangle}_x\|_{\cH_X}^2\\
&\le&2\left(\int_\mcZ y^2d\pi(x,y)+\|f_*\|_\infty^2\right)\times C_X\|\Gamma_*\|_{\textrm{op}}<\infty,
\end{eqnarray*}
and for any $g\in\cH_X$,
\begin{eqnarray*}
\|J_\pi g\|_{\cH_X}&\le&\int_\cX\|(\Gamma_*^{1/2}R^{\langle X\rangle}_x)\otimes(\Gamma_*^{1/2}R^{\langle X\rangle}_x) g\|_{\cH_X}d\pi^X(x)\\
&\le&\int_\cX\|\Gamma_*^{1/2}R^{\langle X\rangle}_x\|_{\cH_X}^2d\pi^X(x)\|g\|_{\cH_X}
\le C_X\|\Gamma_*\|_{\textrm{op}}\|g\|_{\cH_X}.
\end{eqnarray*}
Consequently, $G_\pi\in\cH_X$ and $J_\pi\in\mcL(\cH_X)$.
In fact, $\|J_\pi\|_{\textrm{op}}\le C_X\|\Gamma_*\|_{\textrm{op}}$.
We also observe that $J_\pi$ is self-adjoint semi-positive definite.
Let $T_\pi=(J_\pi+\sigma_\varepsilon^2n^{-1}\textrm{id})^{-1}$, and
\[
H_*(\pi)=f_*+\Gamma_*^{1/2}T_\pi G_\pi,\,\,\pi\in\mcP(\mcZ).
\]
Clearly, $H_*$ is a mapping from $\mcP(\mcZ)$ to $\cH_X$. 
Let $R^{\langle Z\rangle}(z,z')=(1+yy')R^{\langle X\rangle}(x,x')$ and
\begin{eqnarray}
R^{\langle\Pi,X\rangle}(\pi,\pi')&=&\int_{\cX\times\cX}R^{\langle X\rangle}(x,x')d\pi^X(x)d\pi'^X(x'),\nonumber\\
R^{\langle\Pi,Z\rangle}(\pi,\pi')&=&\int_{\mcZ\times\mcZ}R^{\langle Z\rangle}(z,z')d\pi(z)d\pi'(z').\label{eq:RPi}
\end{eqnarray}
Define
\begin{equation}\label{eq:R:Pi}
R^{\langle\Pi\rangle}(\pi,\pi')=\left(1+\frac{R^{\langle\Pi,Z\rangle}(\pi,\pi')}{1-c_\epsilon R^{\langle\Pi,X\rangle}(\pi,\pi')}\right)\cdot\text{id}_{\mathcal{H}_X},\,\,\pi,\pi'\in\mcP(\mcZ),
\end{equation}
where $0<c_\epsilon<C_X^{-1}$ is a scaling constant. By Assumption \ref{A0},
$|R^{\langle\Pi,X\rangle}(\pi,\pi')|\le C_X$ for all $\pi,\pi'\in\mcP(\mcZ)$, hence, such $c_\epsilon$ makes $R^{\langle\Pi\rangle}$ well defined.
Let $\cH_\Pi$ be an RKHS of mappings from $\mcP(\mcZ)$ to $\cH_X$ endowed with reproducing kernel $R^{\langle\Pi\rangle}$.
The following lemma shows that $M_*$ can be explicitly expressed by $H_*$ and $H_*$ is an element of $\cH_\Pi$.
\begin{lemma}\label{lemma:M_*:H_*:GP}
Under Assumption \ref{A0}, (\ref{eq:multiplication:preserving}) and the following series condition
\begin{equation}\label{GP:series:condition}
\sum_{k=1}^\infty\left(\E[\langle W,\varphi^{\langle X\rangle}_k\rangle_{\cH_X}^2]/\rho^{\langle X\rangle}_k\right)^{1/2}<\infty,
\end{equation}
it holds that $H_*\in\cH_\Pi$ and $M_*(\widehat{\pi}_n,\cdot)=H_*(\widehat{\pi}_n)(\cdot)$.    
\end{lemma}    
Note that (\ref{GP:series:condition}) is slightly stronger than $\E[\|W\|_{\cH_X}^2]<\infty$, which is equivalent to 
\[
\sum_{k=1}^\infty\E[\langle W,\varphi^{\langle X\rangle}_k\rangle_{\cH_X}^2]/\rho^{\langle X\rangle}_k<\infty.
\]
The additional strength of (\ref{GP:series:condition}) ensures that the covariance function $C_*$ is sufficiently regular in $\cH_X$ for $H_*$ to belong to $\cH_\Pi$ with a finite norm. 
The kernel $R^{\langle\Pi\rangle}$ in (\ref{eq:R:Pi}) together with (\ref{reproducing:kernel:M}) leads to the construction of $R$ as follows: for $(\pi,x),(\pi',x')\in\mcP(\mcZ)\times\cX$,
\begin{eqnarray*}
R((\pi,x),(\pi',x'))&=&\langle R^{\langle\Pi\rangle}_{\pi} R^{\langle X\rangle}_x,R^{\langle\Pi\rangle}_{\pi'}R^{\langle X\rangle}_{x'}\rangle_{\cH_\Pi}\\
&=&\langle (R^{\langle\Pi\rangle}_{\pi} R^{\langle X\rangle}_x)(\pi'),R^{\langle X\rangle}_{x'}\rangle_{\cH_X}\,\,\,\,(\textrm{by Property\ref{eqn:reproducing:property:ii}})\\
&=&\langle R^{\langle\Pi\rangle}(\pi,\pi')R^{\langle X\rangle}_x,R^{\langle X\rangle}_{x'}\rangle_{\cH_X}\\
&=&\left(1+\frac{R^{\langle\Pi,Z\rangle}(\pi,\pi')}{1-c_\epsilon R^{\langle\Pi,X\rangle}(\pi,\pi')}\right)\langle R^{\langle X\rangle}_x,R^{\langle X\rangle}_{x'}\rangle_{\cH_X}\\
&=&\left(1+\frac{R^{\langle\Pi,Z\rangle}(\pi,\pi')}{1-c_\epsilon R^{\langle\Pi,X\rangle}(\pi,\pi')}\right)R^{\langle X\rangle}(x,x'),
\end{eqnarray*}
where the second last equality holds due to
\[
R^{\langle\Pi\rangle}(\pi,\pi')R^{\langle X\rangle}_x=\left(1+\frac{R^{\langle\Pi,Z\rangle}(\pi,\pi')}{1-c_\epsilon R^{\langle\Pi,X\rangle}(\pi,\pi')}\right)R^{\langle X\rangle}_x.
\]

\section{Appendix: Proofs}\label{sec:appendix}
\subsection{Proof of Lemma \ref{lemma:dimension:reduction}}
Let 
\[
H(z_1,\ldots,z_n,x)=\E[Y_{n+1}| D_n=(z_1,\ldots,z_n),X_{n+1}=x],\,\,z_1,\ldots,z_n\in\mcZ, x\in\cX.
\]
Since $Z_1,\ldots,Z_{n+1}$ is finite exchangeable, we may define 
\[
\widetilde{H}(\{z_1,\ldots,z_n\},x)=H(z_1,\ldots,z_n,x),
\]
which is a measurable mapping that maps an unordered set $\{z_1,\ldots,z_n\}$ to $H(z_1,\ldots,z_n,x)$.
Such definition is meaningful since if two sets $\{z_1,\ldots,z_n\}$ and $\{z_1',\ldots,z_n'\}$ are identical,
there exists a permutation $\tau$ of $1,\ldots,n$ such that $z_i'=z_{\tau(i)}$ for $i=1,\ldots,n$.
By exchangeability, we have
\begin{eqnarray*}
\tilde{H}(\{z_1',\ldots,z_n'\},x)&=&H(z_1',\ldots,z_n',x)=\E[Y_{n+1}| D_n=(z_1',\ldots,z_n'),X_{n+1}=x]\\
&=&\E[Y_{n+1}| D_n=(z_{\tau(1)},\ldots,z_{\tau(n)}),X_{n+1}=x]\\
&=&\E[Y_{n+1}| D_n=(z_1,\ldots,z_n),X_{n+1}=x]=H(z_1,\ldots,z_n,x)\\
&=&\widetilde{H}(\{z_1,\ldots,z_n\},x).
\end{eqnarray*}
Let 
\[
s\left(n^{-1}\sum_{i=1}^n\delta_{z_i}\right)=\{z_1,\ldots,z_n\}
\]
which maps the empirical measure to a unique set of the data points.
Then we have
\[
H(z_1,\ldots,z_n,x)=\widetilde{H}\left(s\left(n^{-1}\sum_{i=1}^n\delta_{z_i}\right),x\right),
\]
which leads to
\[
\E[Y_{n+1}| D_n,X_{n+1}=x]=\widetilde{H}(s(\widehat{\pi}_n),x),
\]
which means $\E[Y_{n+1}| D_n,X_{n+1}=x]$ is in fact a measurable function of $(\widehat{\pi}_n,x)$.
Due to the inclusion of the $\sigma$-fields $\sigma(\widetilde{\pi}_n,X_{n+1})\subseteq\sigma(D_n,X_{n+1})$,
i.e., the information that $\widetilde{\pi}_n$ contains is less than the information that $D_n$ contains,
we have
\begin{eqnarray*}
\E[Y_{n+1}|\widehat{\pi}_n,X_{n+1}]&=&\E[\E[Y_{n+1}| D_n,X_{n+1}]| \widehat{\pi}_n,X_{n+1}]\\
&=&\E[\widetilde{H}(s(\widehat{\pi}_n),X_{n+1})|\widehat{\pi}_n,X_{n+1}]=\widetilde{H}(s(\widehat{\pi}_n),X_{n+1})=\E[Y_{n+1}|D_n,X_{n+1}].
\end{eqnarray*}
Proof is completed.

\subsection{Proofs in Section \ref{sec:asymptotics}}
We calculate the Fr\'{e}chet derivatives of $\ell_{\textbf{n},\textbf{N},B,\lambda}(M)$ as follows:
\begin{eqnarray*}
S_{\textbf{n},\textbf{N},B,\lambda}(M)&\equiv&D\ell_{\textbf{n},\textbf{N},B,\lambda}(M)\\
&=&-\frac{1}{N}\sum_{b=1}^B\sum_{i=1}^{N_b}(Y_{n_b+i}^b-M(\widehat{\pi}_{n_b}^b,X_{n_b+i}^b))K_{(\widehat{\pi}_{n_b}^b,X_{n_b+i}^b)}+P_\lambda M,\\
DS_{\textbf{n},\textbf{N},B,\lambda}(M)\Delta M&=&\frac{1}{N}\sum_{b=1}^B\sum_{i=1}^{N_b}\Delta M(\widehat{\pi}_{n_b}^b,X_{n_b+i}^b)K_{(\widehat{\pi}_{n_b}^b,X_{n_b+i}^b)}+P_\lambda \Delta M,\\
D^2S_{\textbf{n},\textbf{N},B,\lambda}(M)&\equiv&0.
\end{eqnarray*}
By exchangeability, $(\widehat{\pi}_{n_b}^b,Z_{n_b+i}^b)$ and $(\widehat{\pi}_{n_b}^b,Z_{n_b+1}^b)$ are identically distributed. Therefore, the expected Fr\'{e}chet derivatives are 
\begin{eqnarray*}
\E[S_{\textbf{n},\textbf{N},B,\lambda}(M)]&=&\frac{1}{N}\sum_{b=1}^BN_b\E[(M-M_*)(\widehat{\pi}_{n_b}^b,X_{n_b+1}^b)K_{(\widehat{\pi}_{n_b}^b,X_{n_b+1}^b)}]+P_\lambda M\equiv \bar{S}_{\textbf{n},B,\lambda}(M),\\
\E[DS_{\textbf{n},\textbf{N},B,\lambda}(M)\Delta M]&=&\frac{1}{N}\sum_{b=1}^BN_b\E[\Delta M(\widehat{\pi}_{n_b}^b,X_{n_b+1}^b)K_{(\widehat{\pi}_{n_b}^b,X_{n_b+1}^b)}]+P_\lambda\Delta M\equiv D\bar{S}_{\textbf{n},B,\lambda}(M)\Delta M.
\end{eqnarray*}
Obviously, for any $\Delta M,\Delta M'\in\cM$, it holds that
\[
\langle D\bar{S}_{\textbf{n},B,\lambda}(M)\Delta M,\Delta M'\rangle_\cM=\langle\Delta M,\Delta M'\rangle_\cM,
\]
so $D\bar{S}_{\textbf{n},B,\lambda}(M)=\textrm{id}$ is the identity operator on $\cM$. Consequently, 
\begin{eqnarray*}
\bar{S}_{\textbf{n},B,\lambda}(M)&=&\frac{1}{N}\sum_{b=1}^BN_b\E[(M-M_*)(\widehat{\pi}_{n_b}^b,X_{n_b+1}^b)K_{(\widehat{\pi}_{n_b}^b,X_{n_b+1}^b)}]+P_\lambda(M-M_*)+P_\lambda M_*\\
&=&M-M_*+P_\lambda M_*.
\end{eqnarray*}
Let $M_\lambda=(\textrm{id}-P_\lambda)M_*$, so $\bar{S}_{\textbf{n},B,\lambda}(M_\lambda)=0$.
\begin{lemma}\label{boundedness:lemma}
It holds almost surely that
$K_{(\widehat{\pi}_{n_b}^b,X_{n_b+1}^b)}\in\textrm{ker}(\mcC)^\perp$ and
$\|K_{(\widehat{\pi}_{n_b}^b,X_{n_b+1}^b)}\|\le C_\varphi h^{-1/2}$.
\end{lemma}
\begin{proof}[Proof of Lemma \ref{boundedness:lemma}]
Decompose
\[
\textrm{$K_{(\widehat{\pi}_{n_b}^b,X_{n_b+1}^b)}=M_1+M_2$, for $M_1\in\textrm{ker}(\mcC)$ and $M_2\in\textrm{ker}(\mcC)^\perp$.}
\]
Clearly,
\[
\frac{1}{N}\sum_{b=1}^BN_b\E[|M_1(\widehat{\pi}_{n_b}^b,X_{n_b+1}^b)|^2]=V(M_1,M_1)=\langle\mcC M_1,M_1\rangle_\cM=0,
\]
hence, $M_1(\widehat{\pi}_{n_b}^b,X_{n_b+1}^b)=0$ almost surely. By reproducing property, 
\[
\textrm{$\langle M_1, K_{(\widehat{\pi}_{n_b}^b,X_{n_b+1}^b)}\rangle=M_1(\widehat{\pi}_{n_b}^b,X_{n_b+1}^b)=0$, almost surely.}  
\]
Meanwhile,
\begin{eqnarray*}
\langle M_1,M_2\rangle&=&V(M_1,M_2)+\lambda\langle M_1,M_2\rangle_\cM=\langle\mcC M_1,M_2\rangle_\cM+\lambda\langle M_1,M_2\rangle_\cM=0.    
\end{eqnarray*}
So we have $\|M_1\|=0$ almost surely. This shows that $K_{(\widehat{\pi}_{n_b}^b,X_{n_b+1}^b)}=M_2\in\textrm{ker}(\mcC_n)^\perp$ almost surely.
Now consider basis expansion $K_{(\widehat{\pi}_{n_b}^b,X_{n_b+1}^b)}=\sum_{k\ge1}a_k\varphi_k$.
Since 
\begin{equation}\label{eq:K:expansion}
\varphi_k(\widehat{\pi}_{n_b}^b,X_{n_b+1}^b)=\langle\varphi_k,K_{(\widehat{\pi}_{n_b}^b,X_{n_b+1}^b)}\rangle=a_k(1+\lambda\rho_k),
\end{equation}
we have $a_k=\varphi_k(\widehat{\pi}_{n_b}^b,X_{n_b+1}^b)/(1+\lambda\rho_k)$.
By Assumption \ref{A3}, we get
\[
\|K_{(\widehat{\pi}_{n_b}^b,X_{n_b+1}^b)}\|^2=\sum_{k\ge1}\frac{|\varphi_k(\widehat{\pi}_{n_b}^b,X_{n_b+1}^b)|^2}{1+\lambda\rho_k}\le C_\varphi^2 h^{-1}.
\]
Proof is completed.
\end{proof}
\begin{proof}[Proof of Theorem \ref{thm:rate:loo}]
Define
$T(M)=M-S_{\textbf{n},\textbf{N},B,\lambda}(M_\lambda+M)$ for $M\in\cM$.
By proof of Theorem 3.1  in \cite{shang2020colt}, $T(M)$ can be rewritten as
\[
T(M)=-\Gamma(M)-S_{\textbf{n},\textbf{N},B,\lambda}(M_\lambda),   
\]
where
\begin{eqnarray*}
\Gamma(M)&=&\frac{1}{N}\sum_{b=1}^B\sum_{i=1}^{N_b}\left[M(\widehat{\pi}_{n_b}^b,X_{n_b+i}^b)K_{(\widehat{\pi}_{n_b}^b,X_{n_b+i}^b)}-
\E\left[M(\widehat{\pi}_{n_b}^b,X_{n_b+i}^b)K_{(\widehat{\pi}_{n_b}^b,X_{n_b+i}^b)}\right]\right]\\
&=&\frac{1}{N}\sum_{b=1}^B\sum_{i=1}^{N_b}\left[W_i^b(M)-\E\left[W_i^b(M)\right]\right]\\
&=&\frac{1}{N}\sum_{b=1}^B\sum_{i=1}^{N_b}\left[W_i^b(M)-\E\left[W_i^b(M)\big|\boldsymbol{\Pi}\right]\right]
+\frac{1}{N}\sum_{b=1}^B\sum_{i=1}^{N_b}\left[\E\left[W_i^b(M)\big|\boldsymbol{\Pi}\right]-\E\left[W_i^b(M)\right]\right]\\
&\equiv&\Gamma_1(M)+\Gamma_2(M).
\end{eqnarray*}
In the above, we have set $W_i^b(M)\triangleq M(\widehat{\pi}_{n_b}^b,X_{n_b+i}^b)K_{(\widehat{\pi}_{n_b}^b,X_{n_b+i}^b)}$ and $\boldsymbol{\Pi}=\{D_{n_1}^1,\ldots,D_{n_B}^B\}$ with $D_{n_b}^b=(Z_1^b,\ldots,Z_{n_b}^b)$ being the $b$th task data.

Since $W_1^b(M),\ldots,W_{N_b}^b(M)$ are i.i.d. conditional on $\boldsymbol{\Pi}$, and $(W_i^b(M),\widehat{\pi}_{n_b}^b)$ 
is independent of $\{D_{n_{b'}}^{b'}: b'\neq b\}$, we have
\[
\Gamma_2(M)=\frac{1}{N}\sum_{b=1}^B\sum_{i=1}^{N_b}\left[\E\left[W_i^b(M)|D_{n_b}^b\right]-\E[W_i^b(M)]\right]=
\frac{1}{N}\sum_{b=1}^BN_b\left[\E\left[W_1^b(M)|D_{n_b}^b\right]-\E[W_1^b(M)]\right].
\]
Since $D_{n_1}^1,\ldots,D_{n_B}^B$ are independent, and by Lemma \ref{boundedness:lemma} it has $\|W_i^b(M)\|\le C_\varphi h^{-1/2}\|M\|_\infty$, 
it follows by Theorem 3.5 of \cite{pinelis1994} that, for any $M,M'\in\cM$,
\[
\mbbP\left(\|N\sqrt{\frac{h}{\sum_{b=1}^B N_b^2}}[\Gamma_2(M)-\Gamma_2(M')]\|\ge t\right)\le 2e^{-\frac{t^2}{8C_\varphi^2\|M-M'\|_\infty^2}},
\]
and so 
\[
\|N\sqrt{\frac{h}{\sum_{b=1}^B N_b^2}}[\Gamma_2(M)-\Gamma_2(M')]\|_{\psi_2}\le 3C_\varphi\|M-M'\|_\infty,
\] 
where $\|\cdot\|_{\psi_2}$ is the Orlics norm corresponding to $\psi_2(s)\triangleq e^{s^2}-1$.
Combined with Theorem 8.4 of \cite{K08} we have
\[
N\sqrt{\frac{h}{\sum_{b=1}^B N_b^2}}\bigg\|\sup_{M\in\cM_h}\|\Gamma_2(M)\|\bigg\|_{\psi_2}\lesssim a_h:=\int_0^1\psi_2^{-1}(D(\cM_h,\varepsilon,\|\cdot\|_{\infty}))d\varepsilon.
\]
By Lemma 8.1 of \cite{K08}, we have for any $t\ge0$,
\begin{equation}\label{upper:sup:Gamma(f)}
\mbbP\left(\sup_{M\in\cM_h}\|\Gamma_2(M)\|\ge t\right)\le 2e^{-\frac{ht^2}{a_h^2\sum_{b=1}^B N_b^2}}.
\end{equation}
So we have
\[
\sup_{M\in\cM_h}\|\Gamma_2(M)\|=O_\mbbP\left(a_h\sqrt{\frac{\sum_{b=1}^B N_b^2}{N^2h}}\right).
\]
For any $M,M'\in\cH$, it can be seen that $C_\varphi^{-1}h^{1/2}(M-M')/\|M-M'\|\in\cM_h$. So
\begin{eqnarray}
\sup_{\substack{M,M'\in\cM\\ M\neq M'}}\frac{\|\Gamma_2(M)-\Gamma_2(M')\|}{\|M-M'\|}&=&\sup_{\substack{M,M'\in\cM\\ M\neq M'}}
\bigg\|\Gamma_2\left(\frac{C_\varphi^{-1}h^{1/2}(M-M')}{\|M-M'\|}\right)\bigg\|C_\varphi h^{-1/2}\label{eqn:Gamma:2}\\
&\le&\sup_{M\in\cM_h}\|\Gamma_2(M)\| C_\varphi h^{-1/2}=O_\mbbP\left(\frac{a_h\sqrt{\sum_{b=1}^B N_b^2}}{Nh}\right)=o_\mbbP(1), \nonumber
\end{eqnarray}
where the last $o_\mbbP(1)$ follows from condition $a_h\sqrt{\sum_{b=1}^B N_b^2}=o(Nh)$.

To deal with $\Gamma_1$, since conditional $\boldsymbol{\Pi}$, $W_i^b(M)$'s are independent, as well as $\|W_i^b(M-M')\|\le C_\varphi h^{-1/2}\|M-M'\|_\infty$, by Theorem 3.5 of \cite{pinelis1994}, for any $M,M'\in\cM$, we have
\[
\mbbP\left(\|\sqrt{Nh}[\Gamma_1(M)-\Gamma_1(M')]\|\ge t\big|\boldsymbol{\Pi}\right)\le 2e^{-\frac{t^2}{8C_\varphi^2\|M-M'\|_\infty^2}}.
\]
By Theorem 8.4 of \cite{K08}, we have
\[
\bigg\|\sup_{M\in\cM_h}\sqrt{Nh}\|\Gamma_1(M)\|\bigg\|_{\psi_2\big|\boldsymbol{\Pi}}\lesssim a_h:=\int_0^1\psi_2^{-1}(D(\cM_h,\varepsilon,\|\cdot\|_{\infty}))d\varepsilon,
\]
where $\|\cdot\|_{\psi_2\big|\boldsymbol{\Pi}}$ is the Orlicz norm conditional on $\boldsymbol{\Pi}$.
By Lemma 8.1 of \cite{K08}, we have for any $t\ge0$,
\begin{equation}\label{upper:sup:Gamma(f)}
\mbbP\left(\sup_{M\in\cM_h}\|\sqrt{Nh}\Gamma_1(M)\|\ge t\bigg|\boldsymbol{\Pi}\right)\le 2e^{-\frac{t^2}{a_h^2}}.
\end{equation}
Then it holds that
\[
\sup_{M\in\cM_h}\|\Gamma_1(M)\|=O_\mbbP\left(\frac{a_h}{\sqrt{Nh}}\right).
\]
By condition $a_h\sqrt{\sum_{b=1}^B N_b^2}=o(Nh)$, it can be verified that $a_h=o(\sqrt{Nh^2})$. So we have
\begin{equation}\label{eqn:Gamma:1}
\sup_{\substack{M,M'\in\cM\\ M\neq M'}}\frac{\|\Gamma_1(M)-\Gamma_1(M')\|}{\|M-M'\|}=O_{\mbbP}\left(\frac{a_h}{\sqrt{Nh^2}}\right)=o_{\mbbP}(1).
\end{equation}
Combining (\ref{eqn:Gamma:2}) and (\ref{eqn:Gamma:1}), we get
\begin{eqnarray*}
&&\sup_{\substack{M,M'\in\cM\\ M\neq M'}}\frac{\|T(M)-T(M')\|}{\|M-M'\|}\\
&=&\sup_{\substack{M,M'\in\cM\\ M\neq M'}}\frac{\|\Gamma(M)-\Gamma(M')\|}{\|M-M'\|}\\
&\le&
\sup_{\substack{M,M'\in\cM\\ M\neq M'}}\frac{\|\Gamma_1(M)-\Gamma_1(M')\|}{\|M-M'\|}
+\sup_{\substack{M,M'\in\cM\\ M\neq M'}}\frac{\|\Gamma_2(M)-\Gamma_2(M')\|}{\|M-M'\|}=o_{\mbbP}(1).
\end{eqnarray*}
Therefore, with probability approaching one, 
for $M\in\cM$ with $\|M\|\le 2\|T(0)\|=2\|S_{\textbf{n},\textbf{N},B,\lambda}(M_\lambda)\|$,
we get that 
$\|T(M)\|\le 2\|T(0)\|$ and $T$ is a contraction mapping. 
By contraction mapping theorem,
with probability approaching one, there is $\Delta M$ such that $\|\Delta M\|\le 2\|T(0)\|$ and $T(\Delta M)=\Delta M$. This implies that with probability approaching one,
\[
\|\widehat{M}_{\textbf{n},\textbf{N},B,\lambda}-M_*\|\le \|M_\lambda-M_*\|+\|\widehat{M}_{\textbf{n},\textbf{N},B,\lambda}-M_\lambda\|\le\|M_\lambda-M_*\|+2\|S_{\textbf{n},\textbf{N},B,\lambda}(M_\lambda)\|.
\]
We need to bound the second moment of $\|S_{\textbf{n},\textbf{N},B,\lambda}(M_\lambda)\|$.
Recall $e_{n_b,i}^b=Y_{n_b+i}^b-M_*(\widehat{\pi}_{n_b}^b,X_{n_b+i}^b)$.
Since $\E\left[S_{\textbf{n},\textbf{N},B,\lambda}(M_\lambda)\right]=0$, we have
\begin{eqnarray*}
 S_{\textbf{n},\textbf{N},B,\lambda}(M_\lambda)&=&S_{\textbf{n},\textbf{N},B,\lambda}(M_\lambda)-
 \E\left[S_{\textbf{n},\textbf{N},B,\lambda}(M_\lambda)\right]\\
 &=&\frac{1}{N}\sum_{b=1}^B\sum_{i=1}^{N_b}\left[W_i^b(M_\lambda-M_*)-\E[W_i^b(M_\lambda-M_*)]\right]-\frac{1}{N}\sum_{b=1}^B\sum_{i=1}^{N_b} e_{n_b,i}^b K_{(\widehat{\pi}_{n_b}^b,X_{n_b+i}^b)}\\
 &\equiv&\frac{1}{N}\sum_{b=1}^B\sum_{i=1}^{N_b}\left[W_i^b(M_\lambda-M_*)-\E\left[W_i^b(M_\lambda-M_*)\big|D_{n_b}^b\right]\right]\\
 &&+\frac{1}{N}\sum_{b=1}^BN_b\left[\E\left[W_1^b(M_\lambda-M_*)\big|D_{n_b}^b\right]-\E\left[W_1^b(M_\lambda-M_*)\right]\right]\\
 &&-\frac{1}{N}\sum_{b=1}^B\sum_{i=1}^{N_b} e_{n_b,i}^b K_{(\widehat{\pi}_{n_b}^b,X_{n_b+i}^b)}\\
 &\triangleq& I_{1,1}+I_{1,2}-I_2. 
\end{eqnarray*}
We treat the second moments of the three terms respectively.
First, $W_i^b(M_\lambda-M_*), i=1,\ldots,N_b, b=1,\ldots,B$ are mutually independent conditional on $\boldsymbol{\Pi}$. 
By Lemma \ref{boundedness:lemma}, we get that
\begin{eqnarray*}
\E\left[\|I_{1,1}\|^2|\boldsymbol{\Pi}\right]&=&\frac{1}{N^2}\sum_{b=1}^B\sum_{i=1}^{N_b}\left(\E[\|W_i^b(M_\lambda-M_*)\|^2|\boldsymbol{\Pi}]-\|\E[W_i^b(M_\lambda-M_*)|\boldsymbol{\Pi}]\|^2\right)\nonumber\\
&\le&\frac{1}{N^2}\sum_{b=1}^B\sum_{i=1}^{N_b}\E[\|W_i^b(M_\lambda-M_*)\|^2|\boldsymbol{\Pi}]\nonumber\\
&\le&\frac{C_\varphi^2}{N^2h}\sum_{b=1}^BN_b\E[|(P_\lambda M_*)(\widehat{\pi}_{n_b}^b,X_{n_b+1}^b)|^2|\boldsymbol{\Pi}].
\end{eqnarray*}
The last equality follows by the fact that $(\widehat{\pi}_{n_b}^b,X_{n_b+i}^b)$ are identically distributed as $(\widehat{\pi}_{n_b},X_{n_b+1}^b)$
as well as $M_\lambda-M_*=P_\lambda M_*$.
The norm of $P_\lambda M_*$ is controlled by
\begin{equation}\label{eqn:bias}
\|P_\lambda M_*\|=\sup_{\|M\|=1}\langle P_\lambda M_*, M\rangle=\sup_{\|M\|=1}\lambda\langle M_*,M\rangle_\cM\le \sqrt{\lambda}\|M_*\|_\cM,
\end{equation}
Therefore, taking expectation with respect to $\boldsymbol{\Pi}$ and using (\ref{eqn:bias}),
\begin{eqnarray}
 \E[\|I_{1,1}\|^2]&\le&\frac{C_\varphi^2}{N^2h}\sum_{b=1}^BN_b\E[|(P_\lambda M_*)(\widehat{\pi}_{n_b}^b,X_{n_b+1}^b)|^2]=\frac{C_\varphi^2}{Nh}V(P_\lambda M_*,P_\lambda M_*)\nonumber\\
 &\le&\frac{C_\varphi^2}{Nh}\|P_\lambda M_*\|^2\le\frac{C_\varphi^2\|M_*\|_\cM^2\lambda}{Nh}.\label{term:I:1,1}     
\end{eqnarray}
Second, $V_b\triangleq\E\left[W_1^b(M_\lambda-M_*)\big|D_{n_b}^b\right], b=1,\ldots,B$ are independent,
together with Cauchy-Schwartz inequality, we have
\begin{eqnarray}
\E[\|I_{1,2}\|^2]&=&\frac{1}{N^2}\sum_{b=1}^BN_b^2\left[\E[\|V_b\|^2]-\|\E[V_b]\|^2\right]\le\frac{1}{N^2}\sum_{b=1}^BN_b^2 \E[\|V_b\|^2]\nonumber\\
&\le&\frac{1}{N^2}\sum_{b=1}^BN_b^2\E[\|W_1^b(M_\lambda-M_*)\|^2]\nonumber\\
&\le&\frac{C_\varphi^2\max_{1\le b\le B} N_b}{Nh}\times\frac{1}{N}\sum_{b=1}^BN_b\E[|(P_\lambda M_*)(\widehat{\pi}_{n_b}^b,X_{n_b+1}^b)|^2]\nonumber\\
&=&\frac{C_\varphi^2\max_{1\le b\le B} N_b}{Nh}V(P_\lambda M_*,P_\lambda M_*)
\le\frac{C_\varphi^2\|M_*\|_\cM^2\max_{1\le b\le B}N_b\lambda}{Nh}.\label{term:I:1,2}
\end{eqnarray}
Third, we analyze the second moment of $I_2$. Due to independence between the tasks, and $e_{n_b,i}^b$ has zero mean conditional on $D_n^b,X_{n+i}^b$, we get 
\begin{eqnarray}
\E[\|I_2\|^2]&=&\frac{1}{N^2}\sum_{b,b'=1}^B\sum_{i=1}^{N_b}\sum_{i'=1}^{N_{b'}}\E\left[e_{n_b,i}^b e_{n_{b'},i'}^{b'}\langle K_{(\widehat{\pi}_{n_b}^b,X_{n_b+i}^b)},K_{(\widehat{\pi}_{n_{b'}}^{b'},X_{n_{b'}+i'}^{b'})}\rangle\right]\nonumber\\
&=&\frac{1}{N^2}\sum_{b=1}^B\sum_{i,i'=1}^{N_b}\E\left[e_{n_b,i}^b e_{n_b,i'}^b\langle K_{(\widehat{\pi}_{n_b}^b,X_{n_b+i}^b)},K_{(\widehat{\pi}_{n_b}^b,X_{n_b+i'}^b)}\rangle\right].\label{eqn:I_2}
\end{eqnarray}
For $i\neq i'$, $Z_{n_b+i}^b$ and $Z_{n_b+i'}^b$ are i.i.d. conditional on $(D_{n_b}^b,f_b)$. Let $\delta_{n_b,i}^b=e_{n_b,i}^bK_{(\widehat{\pi}_{n_b}^b,X_{n_b+i}^b)}$.
Then we have
\begin{eqnarray*}
\E\left[e_{n_b,i}^b e_{n_b,i'}^b\langle K_{(\widehat{\pi}_{n_b}^b,X_{n_b+i}^b)},K_{(\widehat{\pi}_{n_b}^b,X_{n_b+i'}^b)}\rangle\right]&=&
\E[\E[\langle\delta_{n_b,i}^b,\delta_{n_b,i'}^b\rangle|D_{n_b}^b,f_b]]\\
&=&\E[\langle\E[\delta_{n_b,i}^b|D_{n_b}^b,f_b],\E[\delta_{n_b,i'}^b|D_{n_b}^b,f_b]\rangle]\\
&=&\E[\|\E[\delta_{n_b,i}^b|D_{n_b}^b,f_b]\|^2].
\end{eqnarray*}
Since $Z_{n_b+i}^b$ and $D_{n_b}^b$ are independent conditional on $f_b$, it holds that
\begin{equation}\label{a:smart:equality}
\E[Y_{n_b+i}^b|D_{n_b}^b,f_b,X_{n_b+i}^b]=\E[Y_{n_b+i}^b|f_b,X_{n_b+i}^b]=f_b(X_{n_b+i}^b).
\end{equation}
Also, $M_*(\widehat{\pi}_{n_b}^b,X_{n_b+i}^b)=\E[f_b|\widehat{\pi}_{n_b}^b](X_{n_b+i}^b)$.
Then we have
\begin{eqnarray*}
\E[\delta_{n_b,i}^b|D_{n_b}^b,f_b,X_{n_b+i}^b]&=&\E[e_{n_b,i}^bK_{(\widehat{\pi}_{n_b}^b,X_{n_b+i}^b)}|D_{n_b}^b,f_b,X_{n_b+i}^b]=
\E[e_{n_b,i}^b|D_{n_b}^b,f_b,X_{n_b+i}^b]K_{(\widehat{\pi}_{n_b}^b,X_{n_b+i}^b)}\\
&=&\left(\E[Y_{n_b+i}^b|D_{n_b}^b,f_b,X_{n_b+i}^b]-M_*(\widehat{\pi}_{n_b}^b,X_{n_b+i}^b)\right)K_{(\widehat{\pi}_{n_b}^b,X_{n_b+i}^b)}\\
&=&\left(f_b(X_{n_b+i}^b)-\E[f_b|\widehat{\pi}_{n_b}^b](X_{n_b+i}^b)\right)K_{(\widehat{\pi}_{n_b}^b,X_{n_b+i}^b)},
\end{eqnarray*}
which implies
\[
\E[\delta_{n_b,i}^b|D_{n_b}^b,f_b]=\E[(f_b(X_{n_b+i}^b)-\E[f_b|\widehat{\pi}_{n_b}^b](X_{n_b+i}^b))K_{(\widehat{\pi}_{n_b}^b,X_{n_b+i}^b)}|
D_{n_b}^b,f_b].
\]
By Lemma \ref{boundedness:lemma}, we get
\[
\|\E[\delta_{n_b,i}^b|D_{n_b}^b,f_b]\|\le C_\varphi h^{-1/2}\E[|f_b(X_{n_b+i}^b)-\E[f_b|\widehat{\pi}_{n_b}^b](X_{n_b+i}^b)|\big|D_{n_b}^b,f_b],
\]
hence,
\[
\|\E[\delta_{n_b,i}^b|D_{n_b}^b,f_b]\|^2\le C_\varphi^2 h^{-1}\E[|f_b(X_{n_b+i}^b)-\E[f_b|\widehat{\pi}_{n_b}^b](X_{n_b+i}^b)|^2\big|D_{n_b}^b,f_b].
\]
Therefore,
\[
\E\left[e_{n_b,i}^b e_{n_b,i'}^b\langle K_{(\widehat{\pi}_{n_b}^b,X_{n_b+i}^b)},K_{(\widehat{\pi}_{n_b}^b,X_{n_b+i'}^b)}\rangle\right]
\le C_\varphi^2 h^{-1}\E[|f_b(X_{n_b+i}^b)-\E[f_b|\widehat{\pi}_{n_b}^b](X_{n_b+i}^b)|^2]\equiv C_\varphi^2h^{-1}V_b^2(\textbf{n}),
\]
where recall $V_b^2(\textbf{n})=\E[|f_b(X_{n_b+i}^b)-\E[f_b|\widehat{\pi}_{n_b}^b](X_{n_b+i}^b)|^2]$.
By Assumption \ref{A4} and Lemma \ref{boundedness:lemma}, for $i=i'$,
\begin{eqnarray*}
 \E\left[e_{n_b,i}^b e_{n_b,i'}^b\langle K_{(\widehat{\pi}_{n_b}^b,X_{n_b+i}^b)},K_{(\widehat{\pi}_{n_b}^b,X_{n_b+i'}^b)}\rangle\right]&\le&
 C_\varphi^2 h^{-1}\sigma_e^2.
\end{eqnarray*}
Plugging into (\ref{eqn:I_2}), we get
\begin{equation}\label{term:I:2}
\E[\|I_2\|^2]\le\frac{C_\varphi^2\sum_{b=1}^B N_b^2V_b^2(\textbf{n})}{N^2h}+\frac{C_\varphi^2\sigma_e^2}{Nh}\le\frac{C_\varphi^2V^2(\textbf{n})\max_{1\le b\le B}N_b}{Nh}+\frac{C_\varphi^2\sigma_e^2}{Nh}.
\end{equation}
Recalling $V^2(\textbf{n})=\max_{1\le b\le B}V_b^2(\textbf{n})$.
Plugging (\ref{term:I:1,1}), (\ref{term:I:1,2}) and (\ref{term:I:2}) into $S_{n,B,\lambda}(M_\lambda)$, we get
\begin{eqnarray*}
&&\|S_{\textbf{n},\textbf{N},B,\lambda}(M_\lambda)\|\\
&=&O_\mbbP\left(\lambda^{1/2}(Nh)^{-1/2}+(\lambda\max_{1\le b\le B}N_b)^{1/2}(Nh)^{-1/2}+(Nh)^{-1/2}+V(\textbf{n})(\max_{1\le b\le B}N_b)^{1/2}(Nh)^{-1/2}\right)\\
&=&O_\mbbP\left((\lambda\max_{1\le b\le B}N_b)^{1/2}(Nh)^{-1/2}+(Nh)^{-1/2}+V(\textbf{n})(\max_{1\le b\le B}N_b)^{1/2}(Nh)^{-1/2}\right)\\
&=&O_\mbbP(U(\textbf{n},\textbf{N},B,\lambda)(Nh)^{-1/2}),
\end{eqnarray*}
where $U(\textbf{n},\textbf{N},B,\lambda)=\max\{1,(\lambda\max_{1\le b\le B} N_b)^{1/2},V(\textbf{n})(\max_{1\le b\le B}N_b)^{1/2}\}$.
Together with (\ref{eqn:bias}),
\[
\|\widehat{M}_{\textbf{n},\textbf{N},B,\lambda}-M_*\|\le \|P_\lambda M_*\|+2\|S_{\textbf{n},\textbf{N},B,\lambda}(M_\lambda)\|
=O_\mbbP\left(\lambda^{1/2}+U(\textbf{n},\textbf{N},B,\lambda)(Nh)^{-1/2}\right).
\]
Proof is complete.
\end{proof}
The proof of Lemma \ref{lem:hat:V_n} requires the following Lemma \ref{lem:hat:m} that provides a $L^2$-risk result for estimating $f_b$.
Given task data $D_{n_b}^b$ that are i.i.d. given $f_b$, we estimate $f_b$ as follows:
\begin{eqnarray*}
\widehat{f}_{b,\eta}&=&\argmin_{f\in\cH_X}\ell_{b,\eta}(f)\equiv\argmin_{f\in\cH_X}\left\{\frac{1}{2n_b}\sum_{i=1}^{n_b}|Y_i^b-f(X_i^b)|^2+\frac{\eta}{2}\|f\|_{\cH_X}^2\right\},    
\end{eqnarray*}
where $\eta\asymp n_b^{-\frac{2\gamma}{2\gamma+1}}$ is the smoothing parameter.
To state our estimation result, we make some preparations. 
For $f,f'\in\cH_X$, define 
\[
V_X(f,f')=\E_{X\sim\mu}[f(X)f'(X)]=\int_\cX f(x)f'(x)d\mu(x),
\]
and define a new inner product for $\cH_X$:
\begin{equation}\label{eq:inner:product:cH_X}
\langle f,f'\rangle_X=V_X(f,f')+\eta\langle f,f'\rangle_{\cH_X}.
\end{equation}
Under (\ref{eq:inner:product:cH_X}), $\cH_X$ is also an RKHS with reproducing kernel denoted $K^{\langle X\rangle}:\cX\times\cX\to\mbbR$.
Let $P^{\langle X\rangle}_\eta: \cH_X\to\cH_X$ be a bounded linear operator satisfying 
$\langle P^{\langle X\rangle}_\eta f,f'\rangle_X=\eta\langle f,f'\rangle_{\cH_X}$.
We calculate the Fr\'{e}chet derivatives of $\ell_{b,\lambda}(f)$ as follows:
\begin{eqnarray*}
S_{b,\eta}(f)&\equiv&D\ell_{b,\eta}(f)=-n_b^{-1}\sum_{i=1}^{n_b}(Y_i^b-f(X_i^b))K^{\langle X\rangle}_{X_i^b}+P^{\langle X\rangle}_\eta f,\\
DS_{b,\eta}(f)\Delta f&=&n_b^{-1}\sum_{i=1}^{n_b}\Delta f(X_i^b)K^{\langle X\rangle}_{X_i^b}+P^{\langle X\rangle}_\eta \Delta m,\\
D^2S_{b,\eta}(f)&\equiv&0.
\end{eqnarray*}
The average versions under $f_b$ are 
\begin{eqnarray*}
S_\eta(f)&=&\E_{X\sim\mu}[(f-f_b)(X)K^{\langle X\rangle}_X|f_b]+P^{\langle X\rangle}_\eta f,\\
DS_\eta(f)\Delta f&=&\E_{X\sim\mu}[\Delta f(X)K^{\langle X\rangle}_X]+P^{\langle X\rangle}_\eta \Delta f.
\end{eqnarray*}
\begin{lemma}\label{lem:hat:m}
Suppose that $\varepsilon_i^b$ has finite fourth moment and $R^{\langle X\rangle}$ has decomposition (\ref{mercer:cH_X}), then
 \[
 \E\left[\|\widehat{f}^b_{n,\eta}-f_b\|_X^2|f_b\right]\le C_1(1+\|f_b\|_{\cH_X}^2)n_b^{-\frac{2\beta}{2\beta+1}},
 \]
 where $C_1$ is a constant free of $f_b$. 
\end{lemma}
\begin{proof}[Proof of Lemma \ref{lem:hat:m}]
For the subsequent analysis, we take $\eta\asymp n_b^{-2\beta/(2\beta+1)}$, corresponding to the optimal convergence rate.
For any $f\in\cH_X$,
\[
\langle DS_\eta(f_b)f,f\rangle_X=\E_{X\sim\mu}[f(X)^2]+\eta\|f\|_{\cH_X}^2.
\]
So $DS_\eta(f_b)=\textrm{id}$.
Let $f_{b,\eta}=f_b-P^{\langle X\rangle}_\eta f_b$. Then $S_\eta(f_{b,\eta})=0$.
Define 
\[
\Gamma_X(f)=n_b^{-1}\sum_{i=1}^{n_b}[f(X_i^b)K^{\langle X\rangle}_{X_i^b}-\E_{X\sim\mu}[f(X)K^{\langle X\rangle}_X]].
\]
By boundedness of the eigenfunctions of $R^{\langle X\rangle}$, we have 
\begin{equation}\label{eqn:boundedness:K^X}
\|K^{\langle X\rangle}_x\|_X^2=K^{\langle X\rangle}(x,x)=\sum_{k\ge1}\frac{|\varphi^{\langle X\rangle}_k(x)|^2}{1+\eta\rho^{\langle X\rangle}_k}\le C_{\varphi^{\langle X\rangle}}^2\sum_{k\ge1}\frac{1}{1+\eta\rho^{\langle X\rangle}_k}\equiv C_{\varphi^{\langle X\rangle}}^2g^{-1}.
\end{equation}
Since conditional on $f_b$, $Z_1^b,\ldots,Z_n^b$ are i.i.d., as well as $\|f(X_i^b)K^{\langle X\rangle}_{X_i^b}\|_X\le C_{\varphi^{\langle X\rangle}}g^{-1/2}\|f\|_\infty$, by Theorem 3.5 of \cite{pinelis1994}, for any $f,f'\in\cH_X$, we have
\[
\mbbP\left(\|\sqrt{n_bg}[\Gamma_X(f)-\Gamma_X(f')]\|_X\ge t\right)\le 2e^{-\frac{t^2}{8C_{\varphi^{\langle X\rangle}}^2\|f-f'\|_\infty^2}},
\]
so that 
\[
\|\sqrt{n_bg}\|\Gamma_X(f)-\Gamma_X(f')\|_X\|_{\psi_2}\le 3C_{\varphi^{\langle X\rangle}}\|f-f'\|_\infty.
\]
Let $\cH_g=\{f\in\cH_X: \|f\|_X\le C_{\varphi^{\langle X\rangle}}g^{1/2}\}$ and
\[
a_g\triangleq\int_0^1\psi_2^{-1}(D(\cH_g,\varepsilon,\|\cdot\|_\infty))d\varepsilon.
\]
By similar calculations in Example 4.1 of \cite{shang2025bootstrapnonparametricinferencedata}, we have $g\asymp \eta^{1/(2\beta)}$, hence,
$a_g=O((g/\eta)^{1/(4\beta)})=O(\eta^{-(2\beta-1)/(8\beta^2)})$.
By Theorem 8.4 of \cite{K08}, we have
\[
\bigg\|\sup_{f\in\cH_g}\sqrt{n_bg}\|\Gamma_X(f)\|_X\bigg\|_{\psi_2}\le C a_g,
\]
where $C$ is a universal constant depending on $C_{\varphi^{\langle X\rangle}}$ and our choice of $\psi_2$, hence, is free of $f_b$.
This implies that for any $t\ge0$,
\[
\mbbP\left(\sup_{f\in\cH_g}\sqrt{n_bg}\|\Gamma_X(f)\|_X\ge t\right)\le 2e^{-\frac{t^2}{C^2 a_g^2}}.
\]
Let 
\[
\mcE=\left\{\sup_{f\in\cH_g}\|\Gamma_X(f)\|_X\le \frac{1}{2C_{\varphi^{\langle X\rangle}}}g^{1/2}\right\}.
\]
Since 
$n_bg^2/a_g^2\gtrsim n_b\eta^{(6\beta-1)/(4\beta^2)}\asymp n_b^{(2\beta-1)^2/(2\beta(2\beta+1))}\to\infty$, we have
\[
\mbbP(\mcE^c)\le 2e^{-\frac{n_bg^2}{C^2C_{\varphi^{\langle X\rangle}}^2 a_g^2}}\to0.
\]

Let $\Delta f = \widehat{f}_{b,\eta}-f_{b,\eta}$.
Since
$S_{b,\eta}(\widehat{f}_{b,\eta})=0$, by Taylor expansion and $D^2S_{b,\eta}=0$, we have
\begin{eqnarray*}
0&=&S_{b,\eta}(\widehat{f}_{b,\eta})=S_{b,\eta}(f_{b,\eta})+DS_{b,\eta}(f_{b,\eta})\Delta f\\
&=&S_{b,\eta}(f_{b,\eta})+[DS_{b,\eta}(f_{b,\eta})-DS_\eta(f_{b,\eta})]\Delta f+DS_\eta(f_{b,\eta})\Delta f\\
&=&S_{b,\eta}(f_{b,\eta})+\Gamma_X(\Delta f)+\Delta f.
\end{eqnarray*}
On the event $\mcE$, we have
\begin{eqnarray*}
\|\Gamma_X(\Delta f)\|_X&=&\|\Gamma_X(C_{\varphi^{\langle X\rangle}}^{-1}g^{1/2}\Delta f/\|\Delta f\|_X)\|_X\times 
C_{\varphi^{\langle X\rangle}}g^{-1/2}\|\Delta f\|_X\\
&\le&\sup_{f\in\cH_g}\|\Gamma_X(f)\|_XC_{\varphi^{\langle X\rangle}}g^{-1/2}\|\Delta f\|_X\le \frac{1}{2}\|\Delta f\|_X,
\end{eqnarray*}
which leads to
\begin{equation}\label{eq:delta:m:bound}
\|\widehat{f}_{b,\eta}-f_{b,\eta}\|_X=\|\Delta f\|_X
\le 2\|S_{b,\eta}(f_{b,\eta})\|_X.
\end{equation}
Since $S_\eta(f_{b,\eta})=0$ and $f_{b,\eta}-f_b=-P^{\langle X\rangle}_\eta f_b$ is deterministic given $f_b$, 
and $Z_1^b,\ldots,Z_n^b$ are i.i.d. conditional on $f_b$,
we have
\begin{eqnarray*}
\E\left[\|\widehat{f}_{b,\eta}-f_{b,\eta}\|_X^2I_\mcE|f_b\right]&\le&
4\E[\|S_{b,\eta}(f_{b,\eta})-S_\eta(f_{b,\eta})\|_X^2|f_b]\\    
&=&8\E\left[\|n_b^{-1}\sum_{i=1}^{n_b}\varepsilon_i^b K^{\langle X\rangle}_{X_i^b}\|_X^2\big|f_b\right]
+8\E[\|\Gamma_X(P^{\langle X\rangle}_\eta f_b)\|_X^2|f_b];
\end{eqnarray*}
recalling $\varepsilon_i^b=Y_i^b-f_b(X_i^b)$.
By (\ref{eqn:boundedness:K^X}) and independence of $(\varepsilon_i^b,X_i^b)$ and $f_b$,
\[
\E\left[\|n_b^{-1}\sum_{i=1}^{n_b}\varepsilon_i^b K^{\langle X\rangle}_{X_i^b}\|_X^2\bigg|f_b\right]=
\frac{1}{n_b^2}\sum_{i=1}^{n_b}\E\left[\|\varepsilon_i^b K^{\langle X\rangle}_{X_i^b}\|_X^2\right]
\le\frac{C_{\varphi^{\langle X\rangle}}\sigma_\varepsilon^2}{n_bg},
\]
and by (\ref{eqn:boundedness:K^X}) again,
\begin{eqnarray*}
 &&\E[\|\Gamma_X(P^{\langle X\rangle}_\eta f_b)\|_X^2|f_b]\\
 &=&\frac{1}{n_b^2}\sum_{i=1}^n\left(\E[\|(P^{\langle X\rangle}_\eta f_b)(X_i^b)K^{\langle X\rangle}_{X_i^b}\|_X^2|f_b]-\|\E_{X\sim\mu}[(P^{\langle X\rangle}_\eta f_b)(X)K^{\langle X\rangle}_X|f_b]\|_X^2\right)\\
 &\le&n_b^{-1}\E_{X\sim\mu}[\|(P^{\langle X\rangle}_\eta f_b)(X)K^{\langle X\rangle}_X\|_X^2|f_b]\le\frac{C_{\varphi^{\langle X\rangle}}^2}{n_bg}V_X(P^{\langle X\rangle}_\eta f_b,P^{\langle X\rangle}_\eta f_b)
 \le\frac{C_{\varphi^{\langle X\rangle}}^2}{n_bg}\|P^{\langle X\rangle}_\eta f_b\|_X^2.
\end{eqnarray*}
Obviously,
\[
\|P^{\langle X\rangle}_\eta f_b\|_X=\sup_{\|f\|_X=1}\langle P^{\langle X\rangle}_\eta f_b, f\rangle_X
\le \eta^{1/2}\|f_b\|_{\cH_X},
\]
so we have
\[
 \E[\|\Gamma_X(P^{\langle X\rangle}_\eta f_b)\|_X^2|f_b]\le C_{\varphi^{\langle X\rangle}}^2\|f_b\|_{\cH_X}^2\frac{\eta}{n_bg}.
\]
Therefore,
\[
\E\left[\|\widehat{f}_{b,\eta}-f_{b,\eta}\|_X^2I_\mcE|f_b\right]
\le\frac{8C_{\varphi^{\langle X\rangle}}\sigma_\varepsilon^2}{n_bg}+\frac{8C_{\varphi^{\langle X\rangle}}^2\|f_b\|_{\cH_X}^2\eta}{n_bg}.
\]
Together with 
\[
\|f_{b,\eta}-f_b\|_X^2=\|P^{\langle X\rangle}_\eta f_b\|_X^2\le\|f_b\|_{\cH_X}^2\eta,
\]
we have
\begin{eqnarray}
\E\left[\|\widehat{f}_{b,\eta}-f_b\|_X^2I_\mcE|f_b\right]&\le&2\E\left[\|\widehat{f}_{b,\eta}-f_{b,\eta}\|_X^2I_\mcE|f_b\right]+2\|f_{b,\eta}-f_b\|_X^2\nonumber\\
&\le&\frac{16C_{\varphi^{\langle X\rangle}}\sigma_\varepsilon^2}{n_bg}+\frac{16C_{\varphi^{\langle X\rangle}}^2\|f_b\|_{\cH_X}^2\eta}{n_bg}+2\|f_b\|_{\cH_X}^2\eta\nonumber\\
&\lesssim&(1+\|f_b\|_{\cH_X}^2)n_b^{-\frac{2\beta}{2\beta+1}},
\label{lemma:L2:risk:eqn:1}
\end{eqnarray}
where ``$\lesssim$'' is up to a constant free of $f_b$.
We also need to control $\E\left[\|\widehat{f}_{b,\eta}-f_b\|_X^2I_{\mcE^c}|f_b\right]$. Since
\[
\E\left[\|\widehat{f}_{b,\eta}-f_b\|_X^2I_{\mcE^c}|f_b\right]
\le\left(\E\left[\|\widehat{f}_{b,\eta}-f_b\|_X^4|f_b\right]\right)^{1/2}\mbbP(\mcE^c)^{1/2},
\]
it is sufficient to control $\|\widehat{f}_{b,\eta}-f_b\|_X^4$. 
For $f=\sum_k f_k\varphi^{\langle X\rangle}_k\in\cH_X$, we have
\[
\|f\|_X^2=\sum_k (1+\eta\rho^{\langle X\rangle}_k)f_k^2\le(1/\rho^{\langle X\rangle}_1+\eta)\sum_k f_k^2\rho^{\langle X\rangle}_k=(1/\rho^{\langle X\rangle}_1+\eta)\|f\|_{\cH_X}^2\equiv C_4\|f\|_{\cH_X}^2.
\]
Since
\begin{eqnarray*}
\|\widehat{f}_{b,\eta}-f_b\|_X^4&\le& C_4^2\|\widehat{f}_{b,\eta}-f_b\|_{\cH_X}^4\le C_4^2\left(\|\widehat{f}_{b,\eta}\|_{\cH_X}+\|f_b\|_{\cH_X}\right)^4\\
&\le&16C_4^2\left(\|\widehat{f}_{b,\eta}\|_{\cH_X}^4+\|f_b\|_{\cH_X}^4\right).
\end{eqnarray*}
Since $\ell_{b,\eta}(\widehat{f}_{b,\eta})\le\ell_{b,\eta}(f_b)$, we have
\[
\frac{\eta}{2}\|\widehat{f}_{b,\eta}\|_{\cH_X}^2\le \ell_{b,\eta}(\widehat{f}_{b,\eta})\le\frac{1}{2n_b}\sum_{i=1}^{n_b}|\varepsilon_i^b|^2+\frac{\eta}{2}\|f_b\|_{\cH_X}^2,
\]
together with Cauchy-Schwartz inequality, 
\begin{eqnarray*}
\|\widehat{f}_{b,\eta}\|_{\cH_X}^4&\le&\left(\frac{1}{n_b\eta}\sum_{i=1}^{n_b}|\varepsilon_i^b|^2+\|f_b\|_{\cH_X}^2\right)^2\\
&\le&\frac{2}{n_b^2\eta^2}\left(\sum_{i=1}^{n_b}|\varepsilon_i^b|^2\right)^2+2\|f_b\|_{\cH_X}^4
\le\frac{2}{n_b\eta^2}\sum_{i=1}^{n_b}|\varepsilon_i^b|^4+2\|f_b\|_{\cH_X}^4.
\end{eqnarray*}
So
\[
\|\widehat{f}_{b,\eta}-f_b\|_X^4\le 16C_4^2\left(\frac{2}{n_b\eta^2}\sum_{i=1}^{n_b}|\varepsilon_i^b|^4+3\|f_b\|_{\cH_X}^4\right),
\]
which together with finite fourth moment of $\varepsilon_i^b$, leads to 
\begin{eqnarray*}
\E\left[\|\widehat{f}_{b,\eta}-f_b\|_X^4|f_b\right]&\le&\frac{32C_4^2}{n_b\eta^2}\sum_{i=1}^{n_b}\E[|\varepsilon_i^b|^4|f_b]+48C_4^2\|f_b\|_{\cH_X}^4\\
&\le&32C_4^2\sigma_\varepsilon^4\eta^{-2}+48C_4^2\|f_b\|_{\cH_X}^4.
\end{eqnarray*}
Combining above, we get that, as $n\to\infty$,
\begin{eqnarray}
\E\left[\|\widehat{f}_{b,\eta}-f_b\|_X^2I_{\mcE^c}|f_b\right]&\le&(8C_4\sigma_\varepsilon^2\eta^{-1}+8C_4\|f_b\|_{\cH_X}^2)2^{1/2}
e^{-2^{-1}C^{-2}C_{\varphi^{\langle X\rangle}}^{-2}n_b^{\frac{(2\beta-1)^2}{2\beta(2\beta+1)}}}\nonumber\\
&\le&(1+\|f_b\|_{\cH_X}^2)n_b^{-\frac{2\beta}{2\beta+1}},\label{lemma:L2:risk:eqn:2}
\end{eqnarray}
where the last inequality follows from $\beta>1/2$ hence the exponential term is $o(n_b^{-2\beta/(2\beta+1)})$.
By (\ref{lemma:L2:risk:eqn:1}) and (\ref{lemma:L2:risk:eqn:2}), we get that
\begin{eqnarray*}
\E\left[\|\widehat{f}_{b,\eta}-f_b\|_X^2|f_b\right]&=&\E\left[\|\widehat{f}_{b,\eta}-f_b\|_X^2I_\mcE|f_b\right]+\E\left[\|\widehat{f}_{b,\eta}-f_b\|_X^2I_{\mcE^c}|f_b\right]\\
&\lesssim&(1+\|f_b\|_{\cH_X}^2)n_b^{-\frac{2\beta}{2\beta+1}},
\end{eqnarray*}
where ``$\lesssim$'' is up to a constant free of $f_b$. Proof is completed.
\end{proof}
Next we prove Lemma \ref{lem:hat:V_n}.
\begin{proof}[Proof of Lemma \ref{lem:hat:V_n}]
Due to independence between $X_{n_b+1}^b$ and $(D_{n_b}^b,f_b)$, we have
\[
\E[f_b(X_{n_b+1}^b)|D_{n_b}^b,X_{n_b+1}^b]=\E[f_b|D_{n_b}^b,X_{n_b+1}^b](X_{n_b+1}^b)=\E[f_b|D_{n_b}^b](X_{n_b+1}^b).
\]
Therefore,
\begin{eqnarray*}
\E[|\widehat{f}_{b,\eta}(X_{n_b+1}^b)-f_b(X_{n_b+1}^b)|^2|D_{n_b}^b,X_{n_b+1}^b]    
&=&\E[|f_b(X_{n_b+1}^b)-\E[f_b|D_{n_b}^b](X_{n_b+1}^b)|^2|D_{n_b}^b,X_{n_b+1}^b]\\
&&+\E[|\widehat{f}_{b,\eta}(X_{n_b+1}^b)-\E[f_b|D_{n_b}^b](X_{n_b+1}^b)|^2|D_{n_b}^b,X_{n_b+1}^b],
\end{eqnarray*}
where the equality is due to the vanishing cross term:
\begin{eqnarray*}
&&\E[(f_b(X_{n_b+1}^b)-\E[f_b|D_{n_b}^b](X_{n_b+1}^b))(\widehat{f}_{b,\eta}(X_{n_b+1}^b)-\E[f_b|D_{n_b}^b](X_{n_b+1}^b))|D_{n_b}^b,X_{n_b+1}^b]\\
&=&\E[f_b(X_{n_b+1}^b)-\E[f_b|D_{n_b}^b](X_{n_b+1}^b)|D_{n_b}^b,X_{n_b+1}^b](\widehat{f}_{b,\eta}(X_{n_b+1}^b)-\E[f_b|D_{n_b}^b](X_{n_b+1}^b))=0.
\end{eqnarray*}
So we have
\[
\E[|f_b(X_{n_b+1}^b)-\E[f_b|D_{n_b}^b](X_{n_b+1}^b)|^2|D_{n_b}^b,X_{n_b+1}^b]\le\E[|\widehat{f}_{b,\eta}(X_{n_b+1}^b)-f_b(X_{n_b+1}^b)|^2|D_{n_b}^b,X_{n_b+1}^b].      
\]
Taking expectations w.r.t. $D_{n_b}^b,X_{n_b+1}^b$, we get
\[
V_b(n_b)^2=\E[|f_b(X_{n_b+1}^b)-\E[f_b|D_{n_b}^b](X_{n_b+1}^b)|^2]\le \E[|\widehat{f}_{b,\eta}(X_{n_b+1}^b)-f_b(X_{n_b+1}^b)|^2].
\]
By Lemma \ref{lem:hat:m}, we have
\begin{eqnarray*}
\E[|\widehat{f}_{b,\eta}(X_{n_b+1}^b)-f_b(X_{n_b+1}^b)|^2|D_{n_b}^b,f_b]=V_X(\widehat{f}_{b,\eta}-f_b)\le \|\widehat{f}_{b,\eta}-f_b\|_X^2.
\end{eqnarray*}
Taking expectations w.r.t. $D_{n_b}^b|f_b$ and by Lemma \ref{lem:hat:m}, 
\[
\E[|\widehat{f}_{b,\eta}(X_{n_b+1}^b)-f_b(X_{n_b+1}^b)|^2|f_b]\le\E[\|\widehat{f}_{b,\eta}-f_b\|_X^2|f_b]\le C_1(1+\|f_b\|_{\cH_X}^2)n_b^{-\frac{2\beta}{2\beta+1}}.    
\]
Taking expectations w.r.t. $f_b$,
\[
V_b(n_b)^2\le\E[|\widehat{f}_{b,\eta}(X_{n_b+1}^b)-f_b(X_{n_b+1}^b)|^2]\le C_1(1+\E[\|f_b\|_{\cH_X}^2])n_b^{-\frac{2\beta}{2\beta+1}}.
\]
Proof is completed.
\end{proof}
\begin{proof}[Proof of Theorem \ref{thm:asymp:normality}]
Let $\Delta M=\widehat{M}_{\textbf{n},\textbf{N},B,\lambda}-M_*$.
By Taylor expansion, we have
\begin{eqnarray*}
&&S_{\textbf{n},\textbf{N},B,\lambda}(M_*+\Delta M)-S_{\textbf{n},\textbf{N},B,\lambda}(M_*)-(\bar{S}_{\textbf{n},B,\lambda}(M_*+\Delta M)-\bar{S}_{\textbf{n},B,\lambda}(M_*))\\
&=&DS_{\textbf{n},\textbf{N},B,\lambda}(M_*)\Delta M-D\bar{S}_{\textbf{n},B,\lambda}(M_*)\Delta M\\
&=&\frac{1}{N}\sum_{b=1}^B\sum_{i=1}^{N_b}\left[\Delta M(\widehat{\pi}_{n_b}^b,X_{n_b+i}^b)K_{(\widehat{\pi}_{n_b}^b,X_{n_b+i}^b)}-\E\left[\Delta M(\widehat{\pi}_{n_b}^b,X_{n_b+i}^b)K_{(\widehat{\pi}_{n_b}^b,X_{n_b+i}^b)}\right]\right]\\
&=&\Gamma(\Delta M)\equiv\Gamma_1(\Delta M)+\Gamma_2(\Delta M).
\end{eqnarray*}
By (\ref{eqn:Gamma:2}), (\ref{eqn:Gamma:1}), Theorem \ref{thm:rate:loo} and (\ref{thm:asymp:normality:limit:condition:1}), we have 
\begin{eqnarray*}
 \|\Gamma(\Delta M)\|&\le&\|\Gamma_1(\Delta M)\|+\|\Gamma_2(\Delta M)\|\\
&=&O_\mbbP\left(\left(\frac{a_h\sqrt{\sum_{b=1}^BN_b^2}}{Nh}+\frac{a_h}{\sqrt{Nh^2}}\right)\left(\lambda^{1/2}+(Nh)^{-1/2}\right)\right)\\
&=&O_\mbbP\left(\frac{a_h\sqrt{\sum_{b=1}^BN_b^2}}{Nh}\left(\lambda^{1/2}+(Nh)^{-1/2}\right)\right),
\end{eqnarray*}
in which the last equality holds due to the trivial fact
\[
\frac{a_h\sqrt{\sum_{b=1}^BN_b^2}}{Nh}\ge\frac{a_h}{\sqrt{Nh^2}}.
\]
Since $S_{\textbf{n},\textbf{N},B,\lambda}(M_*+\Delta M)=0$ and $D\bar{S}_{\textbf{n},B,\lambda}(M_*)=\textrm{id}$, we have
\[
\|\widehat{M}_{\textbf{n},\textbf{N},B,\lambda}-M_*+S_{\textbf{n},\textbf{N},B,\lambda}(M_*)\|=O_\mbbP\left(\frac{a_h\sqrt{\sum_{b=1}^BN_b^2}}{Nh}\left(\lambda^{1/2}+(Nh)^{-1/2}\right)\right).
\]
By Lemma \ref{boundedness:lemma} and reproducing property,
\[
\widehat{M}_{\textbf{n},\textbf{N},B,\lambda}(\widehat{\pi},x)-M_*(\widehat{\pi},x)+S_{\textbf{n},\textbf{N},B,\lambda}(M_*)(\widehat{\pi},x)
=O_\mbbP\left(\frac{a_h\sqrt{\sum_{b=1}^BN_b^2}}{Nh^{3/2}}\left(\lambda^{1/2}+(Nh)^{-1/2}\right)\right).
\]
Since
\begin{eqnarray*}
S_{\textbf{n},\textbf{N},B,\lambda}(M_*)(\widehat{\pi},x)&=&-\frac{1}{N}\sum_{b=1}^B\sum_{i=1}^{N_b}e_{n_b,i}^bK_{(\widehat{\pi}_{n_b}^b,X_{n_b+i}^b)}(\widehat{\pi},x)+(P_\lambda M_*)(\widehat{\pi},x)\\
&=&-\frac{1}{N}\sum_{b=1}^B W_b+(P_\lambda M_*)(\widehat{\pi},x),
\end{eqnarray*}
where $W_b=\sum_{i=1}^{N_b}e_{n_b,i}^bK_{(\widehat{\pi}_{n_b}^b,X_{n_b+i}^b)}(\widehat{\pi},x)$.
Due to independence of the tasks, $W_1,\ldots,W_B$ are independent.
Since $\E[e_{n_b,i}^b|\widehat{\pi}_{n_b}^b,X_{n_b+i}^b]=0$, $\E[W_b]=0$.
By CLT, $s^{-1}\sum_{b=1}^B W_b\overset{d}{\to}N(0,1)$, 
where
\[
s^2=\E[|\sum_{b=1}^BW_b|^2]=\sum_{b=1}^B\E[W_b^2].
\]
Next we identify the limit of $s^2$. 
For any $b$,
\begin{eqnarray*}
\E[W_b^2]&=&\sum_{i=1}^{N_b}\E[|e_{n_b,i}^b|^2K((\widehat{\pi}_{n_b}^b,X_{n_b+i}^b),(\widehat{\pi},x))^2]\\
&&+\sum_{i\neq j}\E[e_{n_b,i}^be_{n_b,j}^bK((\widehat{\pi}_{n_b}^b,X_{n_b+i}^b),(\widehat{\pi},x))K((\widehat{\pi}_{n_b}^b,X_{n_b+j}^b),(\widehat{\pi},x))]\equiv \textrm{I}_b+\textrm{II}_b.
\end{eqnarray*}
By (\ref{eq:model:S+N}) and 
\[
e_{n_b,i}^b=Y_{n_b+i}^b-M_*(\widehat{\pi}_{n_b}^b,X_{n_b+i}^b)=\varepsilon_i^b+f_b(X_{n_b+i}^b)-\E[f_b|\widehat{\pi}_{n_b}^b](X_{n_b+i}^b)=\varepsilon_i^b+(f_b-H_*)(X_{n_b+i}^b),
\]
we have
\begin{eqnarray*}
\E[|e_{n_b,i}^b|^2|\widehat{\pi}_{n_b}^b,X_{n_b+i}^b]&=&E[|\varepsilon_i^b+(f_b-H_*)(X_{n_b+i}^b)|^2|\widehat{\pi}_{n_b}^b,X_{n_b+i}^b]\\
&=&\sigma_\varepsilon^2+\E[|(f_b-H_*)(X_{n_b+i}^b)|^2|\widehat{\pi}_{n_b}^b,X_{n_b+i}^b],
\end{eqnarray*}
where the last equality is because
$\E[\varepsilon_i^b(f_b-H_*)(X_{n_b+i}^b)|\widehat{\pi}_{n_b}^b,X_{n_b+i}^b,f_b]=0$ which leads to 
$\E[\varepsilon_i^b(f_b-H_*)(X_{n_b+i}^b)|\widehat{\pi}_{n_b}^b,X_{n_b+i}^b]=0$, so the cross term vanishes.
By Lemma \ref{boundedness:lemma},
$|K((\widehat{\pi}_{n_b}^b,X_{n_b+i}^b),(\widehat{\pi},x))|\le C_\varphi^2 h^{-1}$, we have
\begin{eqnarray*}
\sum_{b=1}^B\textrm{I}_b&=&\sum_{b=1}^B\sum_{i=1}^{N_b}\E[\E[|e_{n_b,i}^b|^2|\widehat{\pi}_{n_b}^b,X_{n_b+i}^b]K((\widehat{\pi}_{n_b}^b,X_{n_b+i}^b),(\widehat{\pi},x))^2]\\
&=&\sum_{b=1}^B\sum_{i=1}^{N_b}\E[(\sigma_\varepsilon^2+\E[|(f_b-H_*)(X_{n_b+i}^b)|^2|\widehat{\pi}_{n_b}^b,X_{n_b+i}^b])K((\widehat{\pi}_{n_b}^b,X_{n_b+i}^b),(\widehat{\pi},x))^2]\\
&=&\sigma_\varepsilon^2\sum_{b=1}^BN_b\E[K((\widehat{\pi}_{n_b}^b,X_{n_b+1}^b),(\widehat{\pi},x))^2]
+\sum_{b=1}^BN_b\E[|(f_b-H_*)(X_{n_b+1}^b)|^2K((\widehat{\pi}_{n_b}^b,X_{n_b+i}^b),(\widehat{\pi},x))^2].
\end{eqnarray*}
By (\ref{eq:K:expansion}), we have
\begin{eqnarray*}
&&\sum_{b=1}^BN_b\E[K((\widehat{\pi}_{n_b}^b,X_{n_b+1}^b),(\widehat{\pi},x))^2]\\
&=&\sum_{b=1}^BN_b\E[(\sum_{k=1}^\infty\frac{\varphi_k(\widehat{\pi}_{n_b}^b,X_{n_b+1}^b)\varphi_k(\widehat{\pi},x)}{1+\lambda\rho_k})^2]\\
&=&\sum_{k,k'=1}^\infty\frac{N}{(1+\lambda\rho_k)(1+\lambda\rho_{k'})}\frac{1}{N}\sum_{b=1}^BN_b\E[\varphi_k(\widehat{\pi}_{n_b}^b,X_{n_b+1}^b)\varphi_{k'}(\widehat{\pi}_{n_b}^b,X_{n_b+1}^b)]\varphi_k(\widehat{\pi},x)\varphi_{k'}(\widehat{\pi},x)\\
&=&\sum_{k,k'=1}^\infty\frac{N}{(1+\lambda\rho_k)(1+\lambda\rho_{k'})}\frac{1}{N}V(\varphi_k,\varphi_{k'})\varphi_k(\widehat{\pi},x)\varphi_{k'}(\widehat{\pi},x)\\
&=&N\sum_{k=1}^\infty\frac{|\varphi_k(\widehat{\pi},x)|^2}{(1+\lambda\rho_k)^2}=Nh^{-1}\sum_{k=1}^\infty\frac{h|\varphi_k(\widehat{\pi},x)|^2}{(1+\lambda\rho_k)^2}.
\end{eqnarray*}
Meanwhile,
\begin{eqnarray*}
\sum_{b=1}^BN_b\E[|(f_b-H_*)(X_{n_b+1}^b)|^2K((\widehat{\pi}_{n_b}^b,X_{n_b+i}^b),(\widehat{\pi},x))^2]&\le&C_\varphi^4h^{-2}\sum_{b=1}^BN_b V_b(n_b)^2. 
\end{eqnarray*}
Therefore,
\begin{equation}\label{thm:asymp:approx:I}
\sum_{b=1}^B\textrm{I}_b=\sigma_\varepsilon^2Nh^{-1}\sum_{k=1}^\infty\frac{h|\varphi_k(\widehat{\pi},x)|^2}{(1+\lambda\rho_k)^2}+
O\left(h^{-2}\sum_{b=1}^BN_b V_b(n_b)^2\right).
\end{equation}
To deal with $\textrm{II}_b$, for $i\neq j$, $\varepsilon_i$ and $\varepsilon_j$ are independent which are also independent of $D_{n_b}^b,X_{n_b+i},X_{n_b+j},f_b$, we have
\begin{eqnarray*}
 &&\E[e_{n_b,i}^be_{n_b,j}^b|\widehat{\pi}_{n_b}^b,X_{n_b+i}^b,X_{n_b+j},f_b]\\   
 &=&\E[(\varepsilon_i^b+(f_b-H_*)(X_{n_b+i}^b))(\varepsilon_j^b+(f_b-H_*)(X_{n_b+j}^b))|\widehat{\pi}_{n_b}^b,X_{n_b+i}^b,X_{n_b+j},f_b]\\
 &=&(f_b-H_*)(X_{n_b+i}^b)(f_b-H_*)(X_{n_b+j}^b).
\end{eqnarray*}
By Lemma \ref{boundedness:lemma},
\begin{eqnarray*}
&&|\E[e_{n_b,i}^be_{n_b,j}^bK((\widehat{\pi}_{n_b}^b,X_{n_b+i}^b),(\widehat{\pi},x))K((\widehat{\pi}_{n_b}^b,X_{n_b+j}^b),(\widehat{\pi},x))]|\\    
&=&|\E[\E[e_{n_b,i}^be_{n_b,j}^b|\widehat{\pi}_{n_b}^b,X_{n_b+i}^b,X_{n_b+j},f_b]K((\widehat{\pi}_{n_b}^b,X_{n_b+i}^b),(\widehat{\pi},x))K((\widehat{\pi}_{n_b}^b,X_{n_b+j}^b),(\widehat{\pi},x))]|\\
&=&|\E[(f_b-H_*)(X_{n_b+i}^b)(f_b-H_*)(X_{n_b+j}^b)K((\widehat{\pi}_{n_b}^b,X_{n_b+i}^b),(\widehat{\pi},x))K((\widehat{\pi}_{n_b}^b,X_{n_b+j}^b),(\widehat{\pi},x))]|\\
&\le&C_\varphi^4h^{-2}\E[|(f_b-H_*)(X_{n_b+i}^b)(f_b-H_*)(X_{n_b+j}^b)|]\le C_\varphi^2h^{-2}V_b(n_b)^2.
\end{eqnarray*}
Therefore, 
\begin{equation}\label{thm:asymp:approx:II}
\sum_{b=1}^B\textrm{II}_b=O\left(h^{-2}\sum_{b=1}^BN_b^2V_b(n_b)^2\right).    
\end{equation}
Combining (\ref{thm:asymp:approx:I}) and (\ref{thm:asymp:approx:II}), and Rate Conditions (\ref{thm:asymp:normality:limit:condition:2}) and (\ref{thm:asymp:normality:limit:condition:5}),
we have
\begin{eqnarray*}
s^2&=&\sum_{b=1}^B\textrm{I}_b+\sum_{b=1}^B\textrm{II}_b
=\sigma_\varepsilon^2Nh^{-1}\sum_{k=1}^\infty\frac{h|\varphi_k(\widehat{\pi},x)|^2}{(1+\lambda\rho_k)^2}+
O\left(h^{-2}\sum_{b=1}^BN_b^2V_b(n_b)^2\right)\\
&=&\sigma_\varepsilon^2\sigma_{(\widehat{\pi},x)}^2Nh^{-1}(1+o(1)).
\end{eqnarray*}
By Conditions (\ref{thm:asymp:normality:limit:condition:3}), (\ref{thm:asymp:normality:limit:condition:4}), we have
\begin{eqnarray*}
&&\frac{\sqrt{Nh}}{\sigma_\varepsilon\sigma_{(\widehat{\pi},x)}}(\widehat{M}_{\textbf{n},\textbf{N},B,\lambda}(\widehat{\pi},x)-M_*(\widehat{\pi},x))\\
&=&s^{-1}\sum_{b=1}^BW_b+Ns^{-1}(P_\lambda M_*)(\widehat{\pi},x)+O_\mbbP\left(\frac{a_h\sqrt{\sum_{b=1}^BN_b^2}}{\sqrt{N}h}\left(\lambda^{1/2}+(Nh)^{-1/2}\right)\right)
\overset{d}{\to}N(0,1).
\end{eqnarray*}
Proof is completed.
\end{proof}
\subsection{Proofs in Section \ref{sec:GP}}
\begin{proof}[Proof of Lemma \ref{lemma:M_*:H_*:GP}]
Let us first prove $H_*\in\cH_\Pi$.
Let $\tau=C_X\|\Gamma_*\|_{\textrm{op}}/2$ and $c=\sigma_\varepsilon^2 n^{-1}+\tau$.
Then $0\preceq J_\pi\preceq 2\tau\textrm{id}$.
So $\|J_\pi-\tau\textrm{id}\|_{\textrm{op}}\le\tau$.
Then the following series holds:
\begin{eqnarray}
\Gamma_*^{1/2}T_\pi G_\pi&=&c^{-1}\Gamma_*^{1/2}\left(\textrm{id}+\frac{J_\pi-\tau\textrm{id}}{c}\right)^{-1}G_\pi=c^{-1}\sum_{l=0}^\infty(-c^{-1})^l\Gamma_*^{1/2}(J_\pi-\tau\textrm{id})^lG_\pi\nonumber\\
&=&c^{-1}\sum_{l=0}^\infty(-c^{-1})^l\sum_{j=0}^l{l\choose j}(-\tau)^{l-j}\Gamma_*^{1/2}(J_\pi)^jG_\pi.\label{eq:GP:decomposition}    
\end{eqnarray}
Let $t_x=\Gamma_*^{1/2}R^{\langle X\rangle}_x$. By direct calculations, when $j\ge1$,
\begin{eqnarray*}
&&\Gamma_*^{1/2}(J_\pi)^jG_\pi\\
&=&\int_\mcZ\int_{\cX^j}(y-f_*(x))\Gamma_*^{1/2}(t_{x_1}\otimes t_{x_1})\cdots(t_{x_j}\otimes t_{x_j})t_xd\pi^X(x_1)\cdots d\pi^X(x_j)d\pi(z)\\
&=&\int_\mcZ\int_{\cX^j}(y-f_*(x))
C_*(x_1,x_2)\cdots C_*(x_{j-1},x_j)C_*(x_j,x)\Gamma_* R^{\langle X\rangle}_{x_1}d\pi^X(x_1)\cdots d\pi^X(x_j)d\pi(z),
\end{eqnarray*}
and 
\[
\Gamma_*^{1/2}(J_\pi)^0G_\pi=\int_\mcZ (y-f_*(x))\Gamma_*R^{\langle X\rangle}_xd\pi(z).
\]
Let $W=\sum_{k=1}^\infty w_k\varphi^{\langle X\rangle}_k$, then
$C_*(x,x')=\sum_{k,l=1}^\infty a_{k,l}\varphi^{\langle X\rangle}_k(x)\varphi^{\langle X\rangle}_l(x')$, where $a_{k,l}=\E[w_kw_l]$.
For $j\ge0$, by direct calculations, it can be checked that
\[
\Gamma_*^{1/2}(J_\pi)^jG_\pi
=\sum_{k_0,l_0,\ldots,k_j,l_j}a_{k_0,l_0}\cdots a_{k_j,l_j}\Phi^X_{l_0,k_1}(\pi)\cdots\Phi^X_{l_{j-1},k_j}(\pi)\Phi^Z_{l_j}(\pi)\varphi^{\langle X\rangle}_{k_0},
\]
where $\psi_{l_{j-1},k_j}(x)=\varphi^{\langle X\rangle}_{l_{j-1}}(x)\varphi^{\langle X\rangle}_{k_j}(x)$, and
\[
\Phi^X_{l_{j-1},k_j}(\pi)=\int_\cX\psi_{l_{j-1},k_j}(x)d\pi^X(x), \Phi^Z_{l_j}(\pi)=\int_{\mcZ}(y-f_*(x))\varphi^{\langle X\rangle}_{l_j}(x)d\pi(x,y).
\]
Let $\cH_Z$ be the RKHS of functions on $\mcZ$ endowed with reproducing kernel $R^{\langle Z\rangle}$ and corresponding inner product $\langle\cdot,\cdot\rangle_{\cH_Z}$. Let
\[
\mcQ^X=\left\{\Phi^X_f: f\in\cH_X\right\},\,\,\mcQ^Z=\left\{\Phi^Z_g: g\in\cH_Z\right\},
\]
where $\Phi^X_f(\pi)=\int_\cX f(x)d\pi^X(x)$ and $\Phi^Z_g(\pi)=\int_\mcZ g(z)d\pi(z)$ for $\pi\in\mcP(\mcZ)$.
Clearly, $\mcQ^X$ and $\mcQ^Z$ are RKHS's with reproducing kernels $R^{\langle\Pi,X\rangle}$ and $R^{\langle\Pi,Z\rangle}$, respectively,
which are defined in (\ref{eq:RPi}).
The corresponding inner products are
\[
\langle\Phi^X_f,\Phi^X_{f'}\rangle_{\mcQ^X}=\langle f,f'\rangle_{\cH_X},\,\,\langle\Phi^Z_g,\Phi^Z_{g'}\rangle_{\mcQ^Z}=\langle g,g'\rangle_{\cH_Z}.
\]
Meanwhile, let $\cH^{\langle j\rangle}_\Pi$ be the RKHS of mappings from $\mcP(\mcZ)$ to $\cH_X$ corresponding to the following kernel
\[
\mcP(\mcZ)\times\mcP(\mcZ)\ni(\pi,\pi')\mapsto [c_\epsilon R^{\langle\Pi,X\rangle}(\pi,\pi')]^jR^{\langle\Pi,Z\rangle}(\pi,\pi')\cdot\textrm{id}_{\cH_X},
\]
which contains the linear span of the following mappings:
\[
\mcP(\mcZ)\ni\pi\mapsto\Phi^X_{f_1}(\pi)\cdots\Phi^X_{f_j}(\pi)\Phi^Z_g(\pi)\varphi^{\langle X\rangle}_k\,\,\,\, f_1,\ldots,f_j\in\cH_X, g\in\cH_Z, k\ge1.
\]
The last assertion is due to the RKHS $\bar{\cH}^{\langle j\rangle}_\Pi$ of functions on $\mcP(\mcZ)$ corresponding to kernel 
\[
\mcP(\mcZ)\times\mcP(\mcZ)\ni(\pi,\pi')\mapsto[c_\epsilon R^{\langle\Pi,X\rangle}(\pi,\pi')]^jR^{\langle\Pi,Z\rangle}(\pi,\pi'),
\]
which contains $[\mcQ^X]^j\mcQ^Z$.
Consequently, the mapping 
$\pi\mapsto\Phi^X_{l_0,k_1}(\pi)\cdots\Phi^X_{l_{j-1},k_j}(\pi)\Phi^Z_{l_j}(\pi)$
belongs to $[\mcQ^X]^j\mcQ^Z$,
we have that the mapping $\pi\mapsto\Phi_{k_0,l_0,\ldots,k_j,l_j}(\pi)\triangleq\Phi^X_{l_0,k_1}(\pi)\cdots\Phi^X_{l_{j-1},k_j}(\pi)\Phi^Z_{l_j}(\pi)\varphi^{\langle X\rangle}_{k_0}$ is a mapping from $\mcP(\mcZ)$ to $\cH_X$
that belongs to $\cH^{\langle j\rangle}_\Pi$.
Then we have
\[
\|\Gamma_*^{1/2}(J_\pi)^jG_\pi\|_{\cH^{\langle j\rangle}_\Pi}\le\sum_{k_0,l_0,\ldots,k_j,l_j}|a_{k_0,l_0}\cdots a_{k_j,l_j}|\times\|\Phi_{k_0,l_0,\ldots,k_j,l_j}\|_{\cH^{\langle j\rangle}_\Pi}.    
\]
Let $\psi_{l_j}(z)=(y-f_*(x))\varphi^{\langle X\rangle}_{l_j}(x)$ which is an element in $\cH_Z$.
In fact, $x\mapsto f_*(x)\varphi^{\langle X\rangle}_{l_j}(x)$ belongs to $\cH_X$ thanks to (\ref{eq:multiplication:preserving});
$(x,y)\mapsto yf_*(x)$ belongs to an RKHS of functions on $\mcZ$ endowed with reproducing kernel $yy'R^{\langle X\rangle}(x,x')$.
It can be seen that
\begin{eqnarray*}
\|\psi_{l_j}\|_{\cH_Z}&\le&\|\varphi^{\langle X\rangle}_{l_j}\|_{\cH_X}+\|f_*\varphi^{\langle X\rangle}_{l_j}\|_{\cH_X}
\le(1+C_d\|f_*\|_{\cH_X})\|\varphi^{\langle X\rangle}_{l_j}\|_{\cH_X}=(1+C_d\|f_*\|_{\cH_X})(\rho^{\langle X\rangle}_{l_j})^{1/2}.
\end{eqnarray*}
By (\ref{eq:multiplication:preserving}) and definition of the $\cH^{\langle j\rangle}_\Pi$ inner product,
\begin{eqnarray*}
&&\|\Phi_{k_0,l_0,\ldots,k_j,l_j}\|_{\cH^{\langle j\rangle}_\Pi}^2\\
&=&c_\epsilon^j\|\Phi^X_{l_0,k_1}\|_{\mcQ^X}^2\cdots\|\Phi^X_{l_{j-1},k_j}\|_{\mcQ^X}^2\|\Phi^Z_{l_j}\|_{\mcQ^Z}^2\|\varphi^{\langle X\rangle}_{k_0}\|_{\cH_X}^2\\    
&=&c_\epsilon^j\|\psi_{l_0,k_1}\|_{\cH_X}^2\ldots\|\psi_{l_{j-1},k_j}\|_{\cH_X}^2\|\psi_{l_j}\|_{\cH_Z}^2\|\varphi^{\langle X\rangle}_{k_0}\|_{\cH_X}^2\\
&\le&c_\epsilon^jC_d^2\|\varphi^{\langle X\rangle}_{l_0}\|_{\cH_X}^2\|\varphi^{\langle X\rangle}_{k_1}\|_{\cH_X}^2\cdots C_d^2\|\varphi^{\langle X\rangle}_{l_{j-1}}\|_{\cH_X}^2\|\varphi^{\langle X\rangle}_{k_j}\|_{\cH_X}^2(1+C_d\|f_*\|_{\cH_X})^2\|\varphi^{\langle X\rangle}_{l_j}\|_{\cH_X}^2\|\varphi^{\langle X\rangle}_{k_0}\|_{\cH_X}^2\\
&=&c_\epsilon^j C_d^{2j}(1+C_d\|f_*\|_{\cH_X})^2\rho^{\langle X\rangle}_{k_0}\rho^{\langle X\rangle}_{l_0}\cdots\rho^{\langle X\rangle}_{k_j}\rho^{\langle X\rangle}_{l_j}.
\end{eqnarray*}
Consequently,
\begin{eqnarray*}
&&\|\Gamma_*^{1/2}(J_\pi)^jG_\pi\|_{\cH^{\langle j\rangle}_\Pi}\\
&\le&c_\epsilon^{j/2}{} C_d^j(1+C_d\|f_*\|_{\cH_X})\sum_{k_0,l_0,\ldots,k_j,l_j}|a_{k_0,l_0}\cdots a_{k_j,l_j}|\times
\left(\rho^{\langle X\rangle}_{k_0}\rho^{\langle X\rangle}_{l_0}\cdots\rho^{\langle X\rangle}_{k_j}\rho^{\langle X\rangle}_{l_j}\right)^{1/2}\\
&=&c_\epsilon^{j/2} C_d^j(1+C_d\|f_*\|_{\cH_X})\left(\sum_{k,l\ge1}|a_{k,l}|(\rho^{\langle X\rangle}_k\rho^{\langle X\rangle}_l)^{1/2}\right)^{j+1}.
\end{eqnarray*}
By (\ref{GP:series:condition}), we have
\[
\sum_{k,l\ge1}|a_{k,l}|(\rho^{\langle X\rangle}_k\rho^{\langle X\rangle}_l)^{1/2}\le\sum_{k,l\ge1}(a_{k,k}a_{l,l})^{1/2}(\rho^{\langle X\rangle}_k\rho^{\langle X\rangle}_l)^{1/2}=\left(\sum_{k\ge1}(a_{k,k}\rho^{\langle X\rangle}_k)^{1/2}\right)^2<\infty.
\]
So $\pi\mapsto\Gamma_*^{1/2}(J_\pi)^jG_\pi$ is a mapping in $\cH^{\langle j\rangle}_\Pi$. By (\ref{eq:GP:decomposition}), this implies that the mapping 
$\pi\mapsto\Gamma_*^{1/2}(J_\pi-\tau\textrm{id})^lG_\pi$ belongs to $\oplus_{j=0}^l\cH^{\langle j\rangle}_\Pi$.
By (\ref{eq:GP:decomposition}) again, the mapping $\pi\mapsto f_*+\Gamma_*^{1/2}T_\pi G_\pi$ belongs to $\cH_\Pi=\cH_X\oplus \left(\oplus_{j=0}^\infty \cH^{\langle j\rangle}_\Pi\right)$. 
The reproducing kernel of $\cH_\Pi$ is 
\begin{eqnarray*}
R^{\langle\Pi\rangle}(\pi,\pi')&=&\text{id}_{\mathcal{H}_X}+\sum_{j=0}^\infty[c_\epsilon R^{\langle\Pi,X\rangle}(\pi,\pi')]^jR^{\langle\Pi,Z\rangle}(\pi,\pi')\cdot\text{id}_{\mathcal{H}_X}\\
&=&\left(1+\frac{R^{\langle\Pi,Z\rangle}(\pi,\pi')}{1-c_\epsilon R^{\langle\Pi,X\rangle}(\pi,\pi')}\right)\cdot\text{id}_{\mathcal{H}_X}
\end{eqnarray*}

Next, let us show that $M_*$ can be explicitly described by $H_*$.
Define
\begin{eqnarray*}
\textbf{f}_*&=&(f_*(X_1),\ldots,f_*(X_n))^\top, \textbf{Y}=(Y_1,\ldots,Y_n)^\top, \\
\textbf{C}_x&=&(C_*(X_1,x),\ldots,C_*(X_n,x))^\top,
\textbf{C}_{n\times n}=[C_*(X_i,X_j)]_{i,j=1}^n. 
\end{eqnarray*}
By (\ref{nonpara:model}), we have
\[
\begin{pmatrix}f(x)\\\mathbf{Y}
\end{pmatrix}\bigg| X_1,\ldots,X_n
\sim N\left(
\begin{pmatrix}f_*(x)\\\mathbf{f}_*
\end{pmatrix},
\begin{pmatrix}
C_*(x,x) & \mathbf{C}_x^\top\\
\mathbf{C}_x & 
\mathbf{C}_{n\times n}+\sigma_\varepsilon^2 \textbf{I}_{n\times n}
\end{pmatrix}
\right).
\]
By conditional normal formula,
\[
f(x)\mid D_n
\sim N\Bigl(
f_*(x)+\mathbf{C}_x^\top
(\mathbf{C}_{n\times n}+\sigma_\varepsilon^2 \textbf{I}_{n\times n})^{-1}
(\mathbf{Y}-\mathbf{f}_*),\;
C_*(x,x)-\mathbf{C}_x^\top
(\mathbf{C}_{n\times n}+\sigma_\varepsilon^2 \textbf{I}_{n\times n})^{-1}
\mathbf{C}_x
\Bigr),
\]
consequently,
\begin{equation}\label{LGM:Mhat}
M_*(\widehat{\pi}_n,x)=\E[f|\widehat{\pi}_n](x)=f_*(x)+\mathbf{C}_x^\top
(\mathbf{C}_{n\times n}+\sigma_\varepsilon^2 \textbf{I}_{n\times n})^{-1}
(\mathbf{Y}-\mathbf{f}_*).
\end{equation}
We will derive an explicit expression of the RHS of (\ref{LGM:Mhat}) in terms of $\widehat{\pi}_n,x$.
Define
\[
A_n\textbf{v}=\sum_{i=1}^n v_i \Gamma_*^{1/2}R^{\langle X\rangle}_{X_i},\,\,\textbf{v}=(v_1,\ldots,v_n)^\top\in\mbbR^n,
\]
which is a linear operator from $\mbbR^n$ to $\cH_X$.
It can be checked that the adjoint of $A_n$ is 
\[
A_n^\dag g=((\Gamma_*^{1/2}g)(X_1),\ldots,(\Gamma_*^{1/2}g)(X_n))^\top,\,\,g\in\cH_X,
\]
which is a linear operator from $\cH_X$ to $\mbbR^n$.
To see this, for any $\textbf{v}=(v_1,\ldots,v_n)^\top\in\mbbR^n$ and $g\in\cH_X$,
\begin{eqnarray*}
\langle\textbf{v},A_n^\dag g\rangle_{\mbbR^n}&=&\langle A_n\textbf{v},g\rangle_{\cH_X}=\sum_{i=1}^nv_i\langle\Gamma_*^{1/2}R^{\langle X\rangle}_{X_i},g\rangle_{\cH_X}\\
&=&\sum_{i=1}^nv_i\langle R^{\langle X\rangle}_{X_i},\Gamma_*^{1/2}g\rangle_{\cH_X}=\sum_{i=1}^nv_i(\Gamma_*^{1/2}g)(X_i)=\textbf{v}^\top((\Gamma_*^{1/2}g)(X_1),\ldots,(\Gamma_*^{1/2}g)(X_n))^\top.
\end{eqnarray*}
It can be checked that 
\begin{equation}\label{eqn:C:and:J}
\textrm{$A_n^\dag A_n=\textbf{C}_{n\times n}$ and $n^{-1}A_nA_n^\dag=J_{\widehat{\pi}_n}$.}
\end{equation}
To see this, for any $\textbf{v}\in\mbbR^n$, by (\ref{eqn:C_*:Gamma_*}) and direct calculations,
\begin{eqnarray*}
 A_n^\dag A_n \textbf{v}&=&A_n^\dag\sum_{i=1}^n v_i\Gamma_*^{1/2}R^{\langle X\rangle}_{X_i}=\sum_{i=1}^nv_iA_n^\dag(\Gamma_*^{1/2}R^{\langle X\rangle}_{X_i})
 \\
 &=&\sum_{i=1}^n v_i\left((\Gamma_*R^{\langle X\rangle}_{X_i})(X_1),\ldots,(\Gamma_*R^{\langle X\rangle}_{X_i})(X_n)\right)^\top\\
&=&\left(\sum_{i=1}^n v_i C_*(X_1,X_i),\ldots,\sum_{i=1}^n v_i C_*(X_n,X_i)\right)^\top=\textbf{C}_{n\times n}\textbf{v},
\end{eqnarray*}
and for any $g\in\cH_X$,
\begin{eqnarray*}
n^{-1}A_nA_n^\dag g&=&n^{-1}A_n ((\Gamma_*^{1/2}g)(X_1),\ldots,\Gamma_*^{1/2}g)(X_n))^\top=n^{-1}\sum_{i=1}^n (\Gamma_*^{1/2}g)(X_i)\Gamma_*^{1/2}R^{\langle X\rangle}_{X_i}\\
&=&n^{-1}\sum_{i=1}^n\langle\Gamma_*^{1/2}g,R^{\langle X\rangle}_{X_i}\rangle_{\cH_X}\Gamma_*^{1/2}R^{\langle X\rangle}_{X_i}=
n^{-1}\sum_{i=1}^n\langle g,\Gamma_*^{1/2}R^{\langle X\rangle}_{X_i}\rangle_{\cH_X}\Gamma_*^{1/2}R^{\langle X\rangle}_{X_i}\\
&=&n^{-1}\sum_{i=1}^n (\Gamma_*^{1/2}R^{\langle X\rangle}_{X_i})\otimes(\Gamma_*^{1/2}R^{\langle X\rangle}_{X_i})g=J_{\widehat{\pi}_n}g.
\end{eqnarray*}
We also notice by (\ref{eqn:C_*:Gamma_*}) that
\[
A_n^\dag(\Gamma_*^{1/2}R^{\langle X\rangle}_x)=((\Gamma_* R^{\langle X\rangle}_x)(X_1),\ldots,(\Gamma_* R^{\langle X\rangle}_x)(X_n))^\top
=(C_*(X_1,x),\ldots,C_*(X_n,x))^\top=\textbf{C}_x,
\]
and
\begin{eqnarray*}
A_n(\textbf{Y}-\textbf{f}_*)&=&\sum_{i=1}^n(Y_i-f_*(X_i))\Gamma_*^{1/2}R^{\langle X\rangle}_{X_i}=nG_{\widehat{\pi}_n}.    
\end{eqnarray*}
Let $T_{\widehat{\pi}_n}=(J_{\widehat{\pi}_n}+\sigma_\varepsilon^2n^{-1}\textrm{id})^{-1}$.
By (\ref{LGM:Mhat}) and (\ref{eqn:C:and:J}), we have
\begin{eqnarray*}
M_*(\widehat{\pi}_n,x)&=&f_*(x)+\langle\textbf{C}_x,(\mathbf{C}_{n\times n}+\sigma_\varepsilon^2 \textbf{I}_{n\times n})^{-1}
(\mathbf{Y}-\mathbf{f}_*)\rangle_{\mbbR^n}\\    
&=&f_*(x)+\langle A_n^\dag(\Gamma_*^{1/2}R^{\langle X\rangle}_x),
(A_n^\dag A_n+\sigma_\varepsilon^2\textbf{I}_{n\times n})^{-1}(\textbf{Y}-\textbf{f}_*)\rangle_{\mbbR^n}\\
&=&f_*(x)+\langle \Gamma_*^{1/2}R^{\langle X\rangle}_x,A_n(A_n^\dag A_n+\sigma_\varepsilon^2\textbf{I}_{n\times n})^{-1}(\textbf{Y}-\textbf{f}_*)\rangle_{\cH_X}\\
&=&f_*(x)+\langle \Gamma_*^{1/2}R^{\langle X\rangle}_x,(A_nA_n^\dag+\sigma_\varepsilon^2\textrm{id})^{-1}A_n(\textbf{Y}-\textbf{f}_*)\rangle_{\cH_X}\\
&=&f_*(x)+\langle \Gamma_*^{1/2}R^{\langle X\rangle}_x,(J_{\widehat{\pi}_n}+\sigma_\varepsilon^2n^{-1}\textrm{id})^{-1}G_{\widehat{\pi}_n}\rangle_{\cH_X}\\
&=&f_*(x)+\langle R^{\langle X\rangle}_x,\Gamma_*^{1/2}T_{\widehat{\pi}_n}G_{\widehat{\pi}_n}\rangle_{\cH_X}=\langle f_*+\Gamma_*^{1/2}T_{\widehat{\pi}_n}G_{\widehat{\pi}_n},R^{\langle X\rangle}_x\rangle_{\cH_X}=\langle H_*(\widehat{\pi}_n),R^{\langle X\rangle}_x\rangle_{\cH_X}.
\end{eqnarray*}
Therefore, $M_*$ is an explicit function of $(\widehat{\pi}_n,x)$.   
Proof is completed.
\end{proof}

\bibliographystyle{apalike}
\bibliography{ref}
\end{document}